%% file: main.tex
\documentclass[sigconf]{acmart}

\AtBeginDocument{%
  \providecommand\BibTeX{{%
    \normalfont B\kern-0.5em{\scshape i\kern-0.25em b}\kern-0.8em\TeX}}}

\usepackage{booktabs}
\usepackage{caption}
\usepackage{hyperref}
\usepackage{multirow}
\usepackage{balance}
\usepackage{enumitem}
\usepackage{framed}
\usepackage{xcolor}
\usepackage{subcaption}

\copyrightyear{2026}
\acmYear{2026}
\setcopyright{cc}
\setcctype{by}
\acmConference[IUI '26]{31st International Conference on Intelligent User Interfaces}{March 23--26, 2026}{Paphos, Cyprus}
\acmBooktitle{31st International Conference on Intelligent User Interfaces (IUI '26), March 23--26, 2026, Paphos, Cyprus}
\acmPrice{}
\acmDOI{10.1145/3742413.3789167}
\acmISBN{979-8-4007-1984-4/2026/03}

\newcommand{\partis}{{participants}}

\newcommand{\insertcmd}{{\texttt{\small insert}}}

\newcommand{\replacecmd}{{\texttt{\small replace}}}

\newcommand{\selectcmd}{{\texttt{\small select}}}
\newcommand{\choosecmd}{{\texttt{\small choose}}}

\newcommand{\sysname}{{VoiceAlign}}

\newcommand{\cmd}{{\texttt{\small cmd}}}
\newcommand{\cmdarg}{{\texttt{\small cmd-arg}}}
\newcommand{\ctx}{{\texttt{\small ctx}}}
\newcommand{\ctxarg}{{\texttt{\small ctx-arg}}}

\begin{document}

\title{VoiceAlign: A Shimming Layer for Enhancing the Usability of Legacy Voice User Interface Systems}

\author{Md Ehtesham-Ul-Haque}
\affiliation{%
  \institution{Pennsylvania State University}
  \city{University Park}
  \state{Pennsylvania}
  \country{USA}
}
\email{mfe5232@psu.edu}

\author{Syed Masum Billah}
\affiliation{%
  \institution{Pennsylvania State University}
  \city{University Park}
  \state{Pennsylvania}
  \country{USA}
}
\email{sbillah@psu.edu}

\input{source/sections/0_abstract}



\begin{teaserfigure}
  \includegraphics[width=0.99\textwidth]{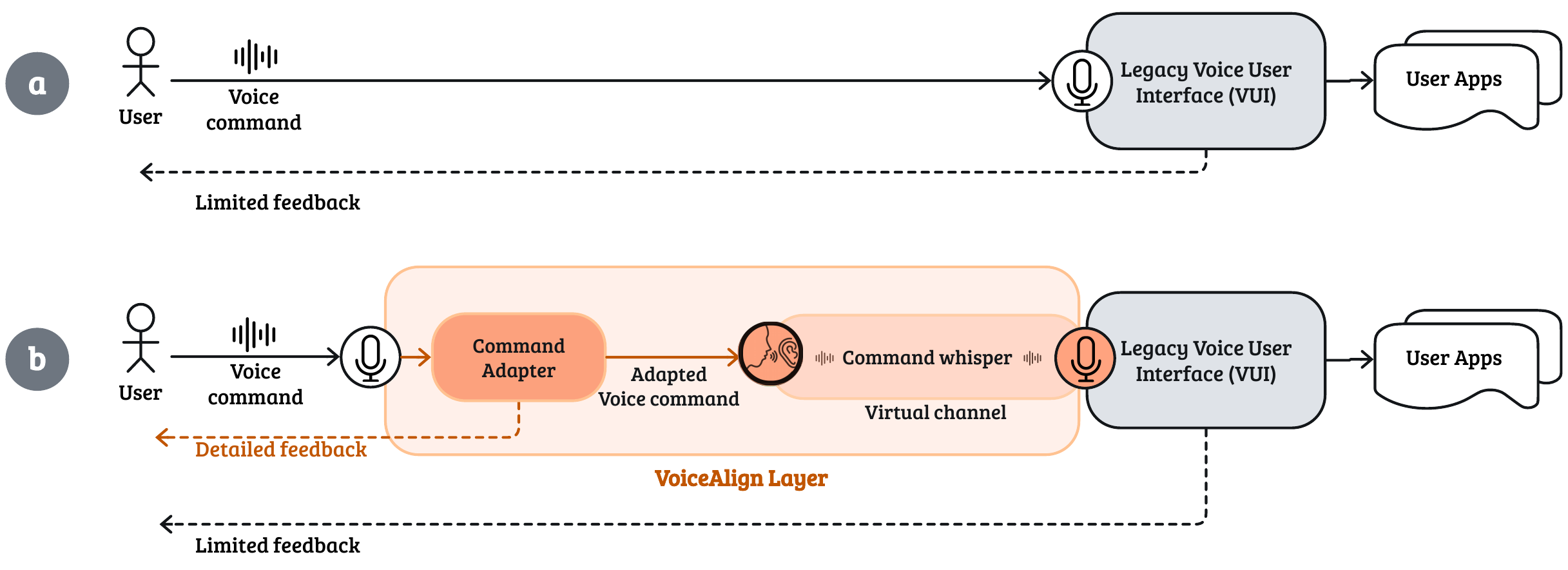}
 \caption{
     An illustration of VoiceAlign's integration with legacy voice user interface (VUI) systems. 
     (a) In a conventional legacy VUI system, users must utter fixed-syntax commands to interact with applications. The VUI interprets these commands and executes them only when they match exact syntactic and semantic requirements, discarding imperfect attempts. 
    (b) VoiceAlign (in orange) functions as a shim layer between system audio input and the legacy VUI. It intercepts user commands, adapts them to match the required syntax and semantics, and relays these corrected commands through a virtual audio channel. This approach frees users from memorizing exact command phrasing and eliminates the need to restate commands during multi-step tasks.}
  \Description{An illustration of VoiceAlign's integration with legacy voice user interface (VUI) systems. 
     (a) In a conventional legacy VUI system, users must utter fixed-syntax commands to interact with applications. The VUI interprets these commands and executes them only when they match exact syntactic and semantic requirements, discarding imperfect attempts. 
    (b) VoiceAlign (in orange) functions as a shim layer between system audio input and the legacy VUI. It intercepts user commands, adapts them to match the required syntax and semantics, and relays these corrected commands through a virtual audio channel. This approach frees users from memorizing exact command phrasing and eliminates the need to restate commands during multi-step tasks.}
  \label{fig:teaser}
\end{teaserfigure}

\maketitle

\input{source/sections/1_intro}
\input{source/sections/2_related_work}

\input{source/sections/3_pilot_study}
\input{source/sections/4_pilot_study_findings}
\input{source/sections/5_system_design}

\input{source/sections/6_findings}

\input{source/sections/7_fine-tuning}
\input{source/sections/8_discussion}
\input{source/sections/9_conclusion}

\begin{acks}
We thank anonymous reviewers for their insightful feedback. This
work was supported in part by NSF Grant \#2326406. The content is solely the responsibility of the authors and does not necessarily represent the official views of the National Science Foundation.
\end{acks}

\balance
\bibliographystyle{ACM-Reference-Format}
\bibliography{Bibliography, Bibliography2, Bibliography3, Bibliography4}

\end{document}

%% file: source/sections/0_abstract.tex
\begin{abstract}

Voice user interfaces (VUIs) are rapidly transitioning from accessibility features to mainstream interaction modalities. Yet most operating systems' built-in voice commands remain underutilized despite possessing robust technical capabilities. Through our analysis of four commercial VUI systems and a formative study with 16 participants, we found that fixed command formats require exact phrasing, restrictive timeout mechanisms discard input during planning pauses, and insufficient feedback hampers multi-step interactions. To address these challenges, we developed VoiceAlign, an adaptive shimming layer that mediates between users and legacy VUI systems. VoiceAlign intercepts natural voice commands, transforms them to match the required syntax using a large language model, and transmits these adapted commands through a virtual audio channel that remains transparent to the underlying system. In our evaluation with 12 participants, VoiceAlign reduced command failures by half, required 25\% fewer commands per task, and significantly lowered cognitive and temporal demands when paired with an existing legacy VUI system. Furthermore, we created a synthetic dataset informed by our studies and fine-tuned a small language model that achieves over 90\% accuracy with 200 ms response time when served locally, eliminating dependence on third-party APIs while enabling real-time interaction on edge devices. This work demonstrates how modern AI techniques can unlock the underutilized potential of legacy VUI systems without requiring system modifications, offering a practical solution without replacing existing infrastructure.

\end{abstract}

%% file: source/sections/1_intro.tex
\section{Introduction}

Voice user interfaces (VUIs) have long served as accessible alternatives to pointing- or touch-based direct manipulation of graphical interfaces~\cite{van1997post}. Historically, users with blindness or upper-limb disabilities relied on VUIs more frequently than their sighted, non-disabled peers~\cite{orcamVoiceActivatedTechnology, dai2003speech}. However, voice input is rapidly shifting from an accessibility feature to a mainstream interaction modality. This transition stems from several converging factors: voice input becoming ubiquitous across devices from smart homes to smart glasses, emerging use cases such as coding where typing complex syntax proves cumbersome, and latency improvements enabling real-time interaction. While VUIs generally prove less efficient than direct manipulation for traditional interaction-oriented activities~\cite{oviatt1997mulitmodal}, they offer unique benefits for hands-free control across diverse environments, including healthcare facilities and augmented reality systems.

This mainstreaming of voice input creates an urgent need to address a persistent paradox: most operating systems include built-in voice command systems that remain underutilized despite their robust technical capabilities. VUI systems like Voice Control~\cite{apple_voice_control}, which enables speech-based interaction as an accessibility feature across Apple's macOS, iOS, and visionOS platforms, possess powerful functionality, including direct access to accessibility APIs, DOM tree structures, and system-level control. Yet these systems require fixed command formats with precise syntax and terminology, creating a fundamental mismatch with natural speech production, which is more verbose, contains disfluencies and false starts, and follows looser organizational structures~\cite{karat1999patterns, lin2024rambler}. Unlike text-based disambiguation, where users can pause indefinitely to review dropdown menus or select from clickable candidate lists, voice commands are ephemeral and temporally constrained --- users cannot visually inspect alternatives before committing, timeout windows prevent leisurely selection processes, and acoustic uncertainty from ASR errors compounds the challenge~\cite{sears2003hands,karat1999patterns}. This fundamental difference imposes cognitive burdens as users must recall exact phrasings in real-time, while minor command variations lead to immediate failures and retries. While modern VUI platforms like Alexa~\cite{ram2018conversational} and ChatGPT desktop~\cite{openai_chatgpt_desktop} have advanced toward natural language understanding~\cite{liu2021benchmarking}, organizations hesitate to replace legacy VUI systems due to prohibitive costs and operational disruptions~\cite{snaplogic2023legacy}. These legacy systems function reliably when users conform to their preconditions, but mastering those preconditions --- speaking in ways contrary to natural speech patterns --- challenges end users~\cite{halverson1999beauty}. A wrapper approach, therefore, offers a more practical solution than complete replacement, as it preserves existing systems while enhancing usability.

To inform our design, we explored the voice command structures across four commercial VUI systems and performed an in-depth analysis of text correction commands supported by all four systems. We then conducted a formative study with 16 participants performing common text correction tasks. Our command structure analysis revealed inconsistencies across command types for target acquisition, target manipulation (e.g., delete a word), and system control. Findings from our formative study showed that participants found the commands restrictive and programming-like, frequently producing minor command variations that VUIs could not support, hampering usability and increasing cognitive demand. Additionally, restrictive timeout mechanisms discarded partial input when users paused to plan multi-component commands, while limited system feedback was insufficient to guide users through error recovery or clarify command requirements. Text correction tasks amplified these challenges because when commands failed due to out-of-vocabulary words, timeouts, or ASR errors, they created error cascades requiring additional corrective commands, representing a worst-case scenario for evaluating VUI usability.

These findings led us to develop \textsc{\textbf{\sysname{}}}, an adaptive shimming layer that mediates between users and legacy VUI systems. As Figure~\ref{fig:teaser} illustrates, \sysname{} consists of two components: a \textit{command adapter} that intercepts the system microphone and uses an LLM to map users' utterances to the correct command syntax, and a \textit{command whisperer} that relays the adapted commands to the legacy VUI through a virtual audio channel. The system provides real-time visual feedback throughout command utterance and employs extended timeout windows to accommodate planning time. 

A summative evaluation with 12 participants demonstrated that \sysname{} reduced command failures by half through suggestions and real-time feedback, and required 25\% fewer commands per task while significantly lowering cognitive load when mediating between the user and a legacy VUI. To enable practical deployment and improve response time, we fine-tuned a 270M parameter language model that achieves over 90\% accuracy in command adaptation with a 200 ms response time when served locally, eliminating dependence on external APIs while enabling real-time interaction on edge devices.

In short, our contributions are:
\begin{itemize}
    \item Analyzed command structures of four legacy VUI systems for consistency patterns and conducted a formative study with 16 participants, identifying command utterance strategies and expectation-structure mismatches for text correction tasks.
    \item Proposed \sysname{}, a shim layer architecture enhancing legacy VUI usability through command adaptation and virtual audio channeling.
    \item Evaluated \sysname{} with 12 participants, demonstrating improved usability without requiring complete system replacement.

    \item Generated a synthetic dataset for command correction informed by study data and fine-tuned a small model capable of running locally and responding in real-time with high accuracy.
\end{itemize}

%% file: source/sections/2_related_work.tex
\section{Background and Related Work}
\label{sec:related-work}

In this section, we describe prior work in voice-/speech-based text correction systems, the correction instructions of the commercially available systems and their challenges, multi-modal correction systems to improve voice commands, and unique challenges in disambiguating voice input.

\subsection{Voice-based Text Correction}
Text correction usually requires users to specify two pieces of information.
First, they need to specify the target location where the correction will occur (target acquisition).
In addition, they need to convey what changes will be performed to correct the text (correction).
In voice-based correction systems, users can acquire the target location through instructions that move the cursor to the target location (navigation-based) or select the target phrase (selection-based).
For correction, users must specify the correction type through a command (e.g., delete, insert, or replace) and additional arguments (e.g., the phrase users want to insert) if required.

Speech-based systems use automated speech recognition (ASR) to capture these dictated commands and arguments.
Two speech-based techniques are commonly used for correction - command-based and re-dictation~\cite{ghosh2020commanding}.
Command-based techniques support a set of fixed-format correction instructions containing a command, a context (e.g., a location), and one or more arguments (e.g., \texttt{\small select <phrase>} and \texttt{\small insert <phrase> before <phrase>}).
These systems utilize a parser to detect valid commands and the arguments to manipulate the text~\cite{ghosh2020commanding}.
Commands are used for both target acquisition and correction.
Corrections can be performed in two steps - using a selection instruction to acquire the target (\texttt{\small select <phrase>}) followed by a correction instruction (\texttt{\small delete that}). 
These two steps can also be combined into a single step (\texttt{\small delete <phrase>}).
Re-dictation replaces the target with the dictated phrase without any predefined commands~\cite{ghosh2020commanding}.
This correction type can also work in either two steps or one step.
Two-step re-dictation also requires target acquisition through a command followed by dictating the correct phrase~\cite{halverson1999beauty}.
In one-step re-dictation, users only dictate the correct phrase, and an underlying algorithm decides the target that aligns the most with the dictation~\cite{mcnair1994improving}.
This is also known as fluid~\cite{vertanen2017ubiquitous}/seamless~\cite{choi2012seamless} correction or correction through re-speaking \cite{sperber2013efficient, vertanen2009automatic}.
Ghosh et al.~\cite{ghosh2018editalk} found that command-based techniques allow more control to users and are preferable for correcting single-word errors, whereas re-dictation approaches are more natural and effective in correcting multi-word errors.

\subsection{Usability Issues of Voice Commands for Text Correction}
Prior work has investigated the challenges of speech-based text correction.
Although speech is faster than keyboard and mouse, correcting errors with dictation is significantly slower~\cite{halverson1999beauty}. 
There are two sources of errors in dictation systems. 
The first is direct/user errors, such as invalid commands or wrong words due to stuttering or pausing. 
The other is indirect/system errors when the ASR fails to detect valid instructions or misrecognizes them due to difficulty separating commands from dictation, noise, or non-native pronunciation~\cite{halverson1999beauty, collings2002usability, sears2003hands, karat1999patterns, ghosh2018editalk, pyae2018investigating}.

Unrecognized or misrecognized voice commands can significantly limit the usability of speech-based text correction. Unrecognized commands confuse users, leading to repetitions (spiral depth)~\cite{pyae2018investigating,sears2003hands,halverson1999beauty} while misrecognition can introduce new errors (cascading effect)~\cite{oviatt1996error,halverson1999beauty,collings2002usability}, making recovery difficult.
Sears et al.~\cite{sears2003hands} found navigation commands misrecognized as dictation, requiring undo operations, and users spent 66\% of their time correcting errors.

Most prior research analyzing voice commands was conducted in the early days of developing ASR and dictation systems. 
They investigated dictation systems with limited features (e.g., using only the \texttt{\small correct} command for making corrections~\cite{halverson1999beauty}) or focused on a limited scope (e.g., investigating only navigation instructions~\cite{sears2003hands}).
In addition, there is a dearth of research investigating the current speech-based text correction systems built on state-of-the-art ASR systems that include many correction commands.
This work aims to fill this gap by comprehensively studying commercial speech-based dictation systems and investigating their usability and challenges.

\subsection{Improving Voice-based Text Correction with Additional Modality}
Although speech is convenient, faster, and more intuitive, it is prone to recognition errors. In addition, proactive control of the microphone is inconvenient and can lead to unintentional command invocation. To mitigate this problem, prior work explored complementing voice commands with additional modalities (e.g., gaze, touch, and gesture), which can lead to effective and robust interaction~\cite{zhao2023heads}.
Multimodal interaction can support natural~\cite{krahnstoever2002real, zhao2022eyesay} and robust experience~\cite{oviatt2000designing, oviatt2000perceptual}, provide flexibility~\cite{oviatt2015paradigm, edwards2002multimodal}, and can reduce cognitive load~\cite{sweller2011cognitive}.
Consequently, researchers explored multimodal systems in various text-based interaction tasks, such as using pen-based gestures to correct ASR errors~\cite{suhm2001multimodal}, combining gesture and speech for typing~\cite{sim2012speak, kristensson2011asynchronous}, using touch to indicate speaking word boundaries for improving ASR performance~\cite{sim2010haptic, oviatt2000taming}, and combining eye gaze and a desktop keyboard for text editing~\cite{sindhwani2019retype}.

Multimodal text correction adopts a two-step dictation-based process, where target acquisition is separated from speech and attributed to a suitable modality (touch, eye gaze, and gesture). Most of the prior work leveraged touch as the modality for target acquisition. For example, Zhao et al.~\cite{zhao2021voice} improved smartphone text correction by using touch for easier target acquisition and speech for correction operations. Touch and speech have also been found to be practical for text modification inside hypothetical automated vehicles.
TouchEditor~\cite{zhan2024toucheditor} used touch gestures (swipe and shape) on a wearable piezoresistive sensor to correct texts on head-mounted devices in speech-unfriendly environments.
Ghosh et al.~\cite{Ghosh2020eyeditor} explored text editing in smart glasses while walking using speech and a hand-held remote.

Researchers have also explored other modalities in mid-air interaction where touch or a keyboard is not available. For example, Talk-and-Gaze~\cite{sengupta2020leveraging} and EyeSayCorrect~\cite{zhao2022eyesay} leveraged eye gaze for target acquisition and speech dictation.
For post-editing machine translation, bimanual gestures and speech have also been explored in prior work using gesture elicitation studies~\cite{herbig2019multi, jamara2021mid}.

While these multimodal systems can make speech-based interaction robust and improve usability, they may not be usable for users who cannot interact with that modality due to impairments. Consequently, making speech-only instructions robust and usable is essential.

\subsection{Challenges of Disambiguating Voice Input}

While disambiguation interfaces have been extensively studied in text-based contexts—including dropdown refinement~\cite{hearst2009search}, query suggestions~\cite{amershi2019guidelines}, and candidate selection interfaces—voice command correction presents fundamentally distinct challenges. Unlike text input, where users can pause indefinitely, visually inspect their typed query, and iteratively refine character-by-character, voice commands are ephemeral and temporally constrained --- users must articulate complete multi-component commands within strict timeout windows (typically 1-2 seconds between components) or lose all progress~\cite{sears2003hands}. This temporal pressure is compounded by acoustic uncertainty, as ASR errors, homophones, and pronunciation variations introduce ambiguity before semantic interpretation begins—a layer of uncertainty absent in text interfaces where input fidelity is guaranteed~\cite{karat1999patterns}. 

Moreover, the spontaneous nature of speech production itself complicates disambiguation. Unlike text, where users can draft, review, and edit before submission, voice commands emerge in real-time with disfluencies, false starts, and self-corrections that are natural artifacts of unplanned speech~\cite{clark1996using, lin2024rambler}. Users cannot reliably produce perfectly formed commands without advance planning. This challenge is further exacerbated by the difficulty of maintaining command mode boundaries --- users must continuously manage whether they are issuing commands or thinking aloud, as systems cannot reliably distinguish intentional commands from planning utterances, self-talk, or environmental noises~\cite{oviatt2000designing, liu2022typist}.

Our work addresses these voice-specific challenges through a shimming layer that preprocesses commands before passing them to the legacy VUIs. We provide extended timeout windows to accommodate spontaneous speech planning, use LLM-based parsing to separate noise and disfluencies from command components and map them to correct syntax, and generate clarifying questions for incomplete commands rather than discarding partial input.

\section{Analyzing Commands in Legacy VUI Systems}
\label{sec:analysis}

To better understand VUI commands, their structures, and their applications in personal computer use, we conducted a systematic analysis of their command structures. We first identified VUI systems with fixed-format commands that met four criteria: (i) fully-featured with comprehensive command sets, (ii) widely available to users, (iii) commercially popular with substantial user bases, and (iv) supporting interaction across common operating systems. This selection process yielded four systems: Voice Control (Apple)~\cite{apple_voice_control}, Dragon Speech Recognition (Nuance)~\cite{nuance_dragon}, Voice Access for Windows (Microsoft)~\cite{microsoft_voice_access}, and Voice Access for Android (Google)~\cite{google_voice_access}. We then analyzed the command capabilities of all four systems and identified that dictation and text correction commands were universally supported. Therefore, we focused on text correction tasks with voice commands (Table~\ref{tab:voice_commands}), which prove especially challenging with voice~\cite{halverson1999beauty, sears2003hands, collings2002usability}.

Next, each author independently analyzed the commands supported by each system, using them to perform common text correction tasks (e.g., selecting, deleting, inserting, replacing, and correcting text) to understand in depth how each command functioned. During this analysis, the authors sought to identify command structures, their intricacies and components, what factors made commands successful or caused them to be discarded by the system, and the feedback provided during interaction. Each author took detailed notes independently and created command templates based on their analysis. Following individual analysis, the authors met to discuss their notes, identify common themes regarding text correction commands use, and consolidate their templates into a unified, fully-quantified command template that specifies valid component combinations. We discuss this template, its components, valid combinations, and command utterance techniques in detail in this section.

\begin{table*}[t!]
    \scriptsize
    \centering
    \begin{tabular}{l|p{3cm}|p{3cm}|p{3cm}|p{3cm}}
         \textbf{Operation} & \textbf{Voice Control (Apple)} & \textbf{Dragon Speech (Nuance)} & \textbf{Voice Access for Windows (Microsoft)} & \textbf{Voice Access for Android (Google)}\\\toprule
         \multirow{4}{*}{Mode Switching}    & $\bullet$ \texttt{dictation mode} & $\bullet$ \texttt{switch to dictation mode} & $\bullet$ \texttt{default mode}         &    \\
                                            & $\bullet$ \texttt{command mode}   & $\bullet$ \texttt{switch to command mode} & $\bullet$ \texttt{commands mode}    &  \\
                                            & $\bullet$ \texttt{spelling mode}   & $\bullet$ \texttt{switch to spelling mode} & $\bullet$ \texttt{dictation mode}    &  \\\midrule
          \multirow{3}{*}{Phrase Dictation} & $\bullet$ \texttt{<phrase>}         & $\bullet$ \texttt{<phrase>} & $\bullet$ \texttt{<phrase>}         & $\bullet$ \texttt{<phrase>}    \\
                                   & $\bullet$ \texttt{type <phrase>}    &            & $\bullet$ \texttt{type <phrase>}    & $\bullet$ \texttt{type <phrase>} \\
                                   &                    &            & $\bullet$ \texttt{dictate <phrase>} &               \\\midrule
        \multirow{3}{*}{Navigation} & $\bullet$ \texttt{move before <phrase>}         & $\bullet$ \texttt{move before <n> characters} & $\bullet$ \texttt{move before <phrase>} &  $\bullet$ \texttt{move before <phrase>}    \\
                                   & $\bullet$ \texttt{move after <phrase>}    & $\bullet$ \texttt{move down <n> lines}           & $\bullet$ \texttt{move after <phrase>}    & $\bullet$ \texttt{move after <phrase>} \\
                                   &                           &                               &                         &  $\bullet$ \texttt{move between <phrase> and <phrase>}  \\\midrule
         \multirow{4}{*}{Phrase Selection} & $\bullet$ \texttt{select <phrase>}        & $\bullet$ \texttt{select <phrase>}                    & $\bullet$ \texttt{select <phrase>}        & $\bullet$ \texttt{select <phrase>}\\
                                    & $\bullet$ \texttt{select previous word}   & $\bullet$ \texttt{select <phrase> through <phrase>}   & $\bullet$ \texttt{select previous word}   & $\bullet$ \texttt{select from <phrase> to <phrase>} \\
                                    & $\bullet$ \texttt{select next word}       & $\bullet$ \texttt{select next <n> words}              & $\bullet$ \texttt{select next word}       & \\\midrule
          \multirow{2}{*}{Number Selection} & $\bullet$ \texttt{choose <number>}    & $\bullet$ \texttt{choose <number>}    & $\bullet$ \texttt{click <number>} & \\
                                            & $\bullet$ \texttt{<number>}           & $\bullet$ \texttt{<number>}           & $\bullet$ \texttt{<number>}                & \\\midrule
         \multirow{6}{*}{Insertion} & $\bullet$ \texttt{insert <phrase> before <phrase>}    & $\bullet$ \texttt{insert before <phrase>} $\rightarrow$ Phrase Dictation & $\bullet$ \texttt{insert before <phrase>} $\rightarrow$ Phrase Dictation & $\bullet$ \texttt{Insert <phrase> before <phrase>} \\
                                    & $\bullet$ \texttt{insert <phrase> after <phrase>}     & $\bullet$ Cursor Movement $\rightarrow$ Phrase Dictation                 & $\bullet$ \texttt{insert after <phrase>} $\rightarrow$ Phrase Dictation & $\bullet$ \texttt{insert <phrase> after <phrase>} \\
                                    & $\bullet$ Cursor Movement $\rightarrow$ Phrase Dictation                 &                                               & $\bullet$ Cursor Movement $\rightarrow$ Phrase Dictation & $\bullet$ \texttt{insert <phrase> between <phrase> and <phrase>} \\\midrule
          \multirow{7}{*}{Deletion} & $\bullet$ \texttt{delete <phrase>}                        & $\bullet$ \texttt{delete <phrase>} & $\bullet$ \texttt{delete <phrase>}        & $\bullet$ \texttt{delete <phrase>} \\
                                    & $\bullet$ Phrase Selection $\rightarrow$ \texttt{delete selection}    & $\bullet$ \texttt{delete from <phrase> to <phrase>}           & $\bullet$ Phrase Selection $\rightarrow$ \texttt{delete that} & $\bullet$ \texttt{delete from <phrase> to <phrase>} \\
                                    & $\bullet$ Phrase Selection $\rightarrow$ \texttt{delete that}         & $\bullet$ Cursor Movement $\rightarrow$ \texttt{delete last <n> words}    & $\bullet$ Phrase Selection $\rightarrow$ \texttt{scratch that} & $\bullet$ Phrase Selection $\rightarrow$ \texttt{delete selected text} \\
                                    &                                                   & $\bullet$ Cursor Movement $\rightarrow$ \texttt{backspace <n>}            & $\bullet$ Phrase Selection $\rightarrow$ \texttt{strike that} & \\\midrule
         \multirow{4}{*}{Replacement}   & $\bullet$ \texttt{replace <phrase> with <phrase>}  & Phrase Selection $\rightarrow$ Phrase Dictation & Phrase Selection $\rightarrow$ Phrase Dictation  & $\bullet$ \texttt{replace <phrase> with <phrase>} \\
                                        & $\bullet$ \texttt{change <phrase> to <phrase>}     &                              &                               & $\bullet$ \texttt{replace everything between <phrase> and <phrase> with <phrase>} \\
                                        & $\bullet$ Phrase Selection $\rightarrow$ Phrase Dictation             &                              &                               &        \\\midrule
         \multirow{2}{*}{Fixing} & $\bullet$ \texttt{correct <phrase>} & $\bullet$ \texttt{correct <phrase>} & $\bullet$ \texttt{correct <phrase>} & \\
          & $\bullet$ Phrase Selection $\rightarrow$ \texttt{correct that} & $\bullet$ Phrase Selection $\rightarrow$ \texttt{correct that} & $\bullet$ Phrase Selection $\rightarrow$ \texttt{correct that} &  \\\bottomrule
    \end{tabular}
    \caption{A sampler of commands supported by four commercial VUI systems.}
    \Description{A sampler of commands supported by four commercial VUI systems.}
    \label{tab:voice_commands}
\end{table*}

\begin{figure*}[t!]
    \centering    
    \includegraphics[width=.85\textwidth]{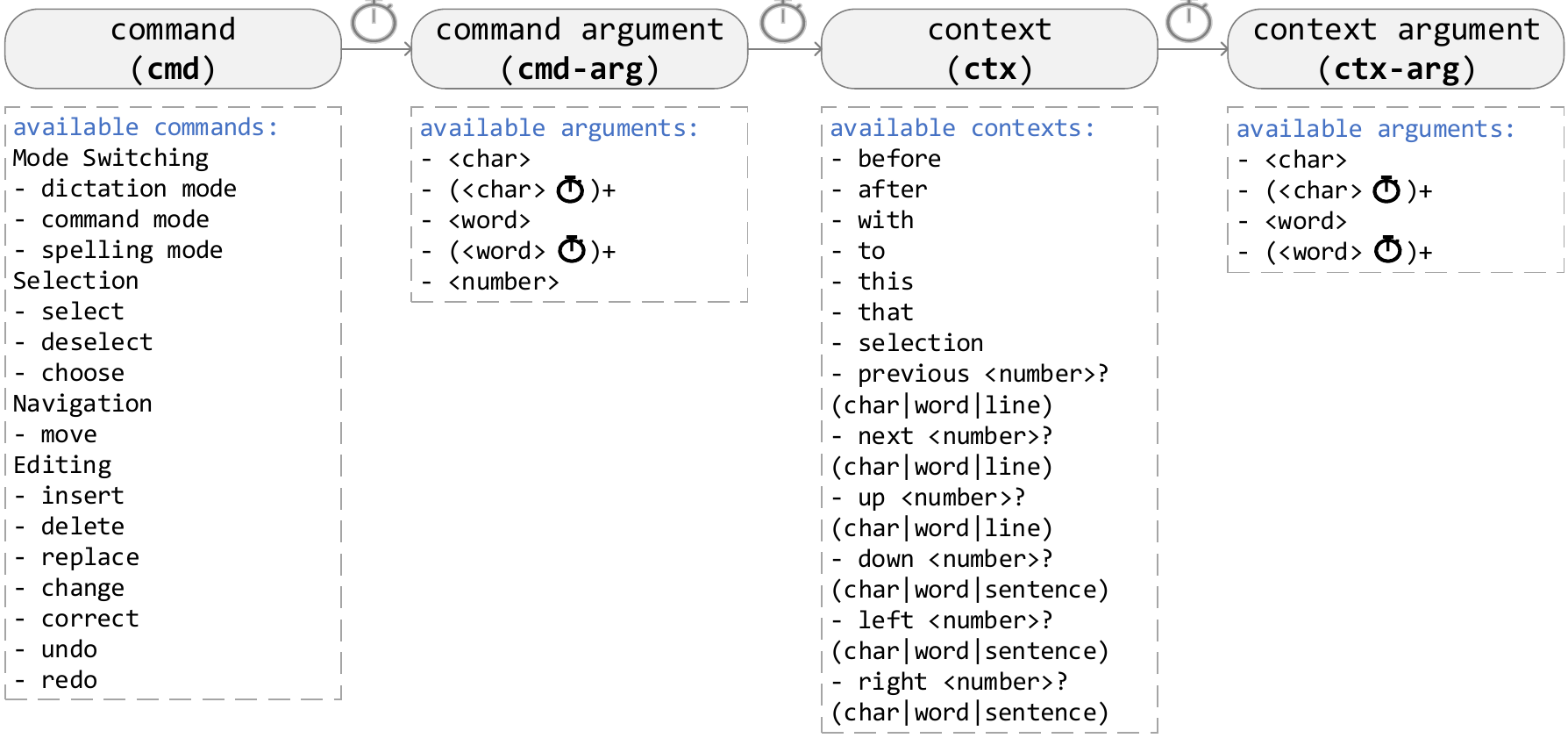}
    \caption{
    The fully quantified command template with four components and a sampler of available values for each component. An icon indicates the timer that specifies the threshold between two utterances to be considered part of the same instruction. Note that the timer applies to both inter- and intra-component utterances.}
    \Description{The fully quantified command template with four components and a sampler of available values for each component. An icon indicates the timer that specifies the threshold between two utterances to be considered part of the same instruction. Note that the timer applies to both inter- and intra-component utterances.}
	\label{fig:cmnd-template}
\end{figure*}

\begin{figure*}[t]
    \centering
    \includegraphics[width=.85\textwidth]{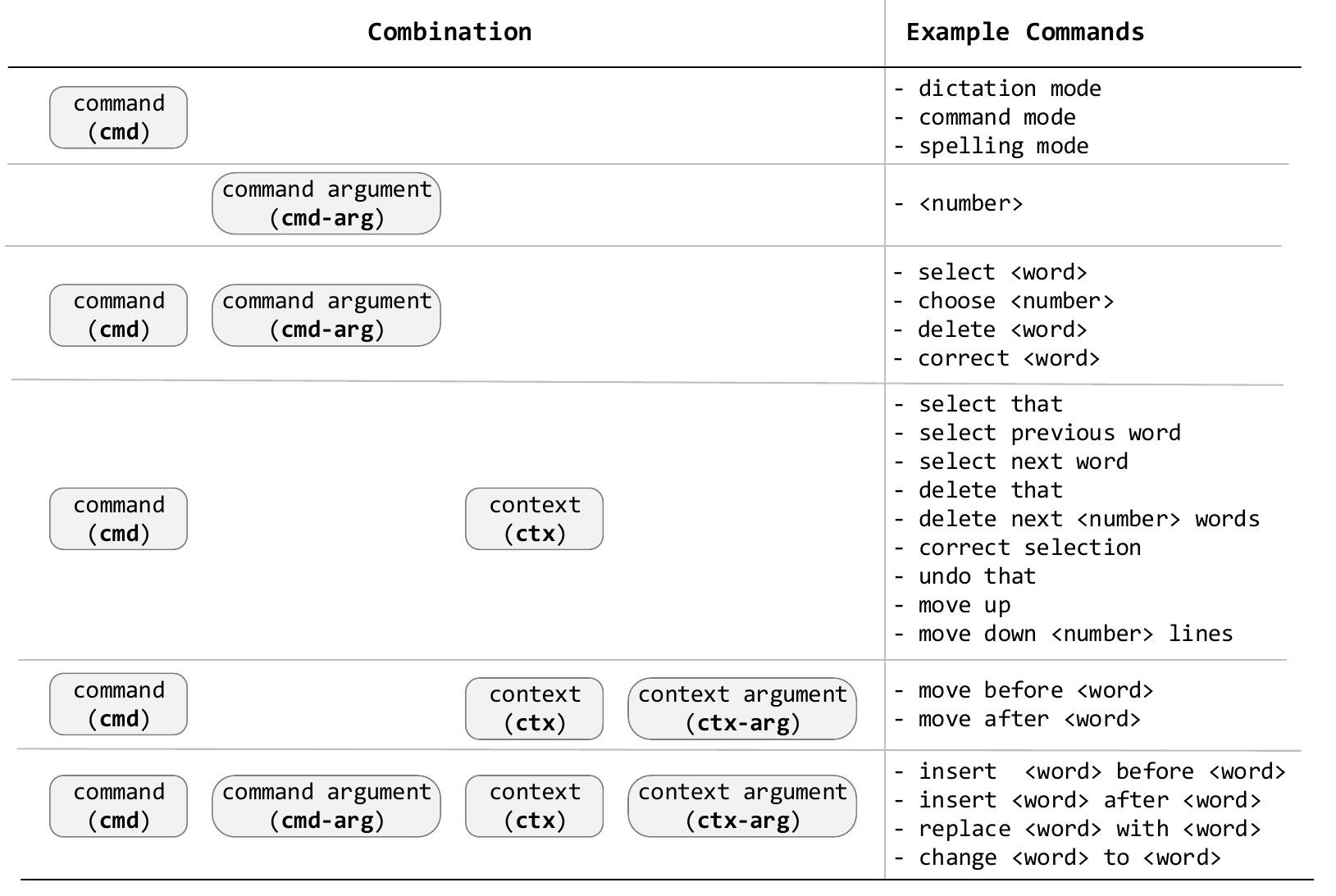}
    \caption{
        Six valid combinations of the four components with example commands for each combination.}
    \Description{Six valid combinations of the four components with example commands for each combination.}
	\label{fig:cmnd-comb}
\end{figure*}

\paragraph{Fully Quantified Command Template}
Legacy VUI systems require users to memorize and execute precise command structures with fixed syntax.
These systems typically organize commands into distinct operational categories (Table~\ref{tab:voice_commands}): mode switching, phrase dictation, navigation, selection, and manipulation (insertion, deletion, replacement).
We generalize their command structures into four components, as shown in Figure~\ref{fig:cmnd-template}: \texttt{[command (cmd)] <command argument (cmd-arg)> [context (ctx)] <context argument (ctx-arg)>}.

Using this template, we identified six common combinations (Fig.~\ref{fig:cmnd-comb}) that use various subsets of these four components.
While some command structures remain consistent across all four systems, significant variations exist, especially in complex operations. For example, replacement operations use different syntactic structures:
These variations highlight the irregularity, implicit complexity, and ad hoc nature of command structures in legacy VUI systems.

\paragraph{Command Complexity}
For any command to succeed, users must correctly populate all required components with valid information. We rank the complexity of commands based on their component combinations:

\begin{itemize}
    \item \textit{Single-component commands} like mode switching impose minimal cognitive load.
    \item \textit{Simple commands} for selection, deletion, or correction typically require two components: a command keyword and a parameter.
    \item \textit{Complex commands} for insertion or replacement require all four components (e.g., \texttt{insert <word> before <word>}), creating maximal cognitive load by requiring users to organize multiple pieces of information in the correct sequence.
    \item \textit{Special cases} exist, such as the "\texttt{\small choose}" command, where the first component (command keyword) becomes optional, allowing users to select by simply speaking a number.
\end{itemize}

\paragraph{Timeout Mechanisms and Their Impact}
A critical usability constraint in legacy VUI systems is the timeout mechanism. As shown in Figure~\ref{fig:cmnd-template}, each transition between components has an associated timeout threshold (inter-component; 1 to 2 seconds). This timeout also applies to an individual component with multiple parts (intra-component; e.g., a phrase as a \texttt{\small cmd-arg}). If users pause longer than the allowed threshold, the system discards the partial command and often interprets subsequent speech as new input.

%% file: source/sections/3_pilot_study.tex
\section{Formative User Study}
\label{sec:pilot}

We conducted an IRB-approved study with 16 \partis{} (14 males, 2 females, ages 24-35) recruited through university mailing lists and word of mouth. All participants were fluent English speakers with varying accents, without speech difficulties, and had prior experience with voice interfaces such as Siri, Google Home, or Alexa. We refer to participants as P1 through P16 throughout our analysis.

\subsection{Study Design}
Our study comprised two complementary parts. In the \textit{first} observational part, participants used voice commands to correct erroneous sentences while we documented their correction processes, instruction choices, target acquisition strategies, and reactions to the VUI system. We designed four representative text correction tasks based on common error patterns identified in Palin et al.'s~\cite{palin2019people} mobile typing dataset:

\begin{itemize}
    \item \textbf{T1: } Text correction by \textit{inserting a phrase}
    \item[] (e.g., The enforcement has responsibility ... $\rightarrow$ The \textbf{\textit{law}} enforcement has responsibility...)
    \item \textbf{T2: } Text correction by \textit{deleting a phrase}
    \item[] (e.g., Was \textbf{\textit{is}} it a car wreck? $\rightarrow$ Was it a car wreck?)
    \item \textbf{T3: } Text correction by \textit{replacing a phrase} 
    \item[] (e.g., Every employee had an \textbf{\textit{internet}} email address. $\rightarrow$ Every employee had an \textbf{\textit{internal}} email address.)
    \item \textbf{T4: } Text correction by \textit{fixing the typo of a phrase}
    \item[] (e.g., There were \textbf{\textit{freqwuent}} electricity and water shortages. $\rightarrow$ There were \textbf{\textit{frequent}} electricity and water shortages.)
\end{itemize}

Each participant completed 5 trials per task type, totaling 20 trials presented in randomized order. In the \textit{second} part, we conducted semi-structured interviews to explore participants' thought processes when formulating instructions, their strategies for task completion and error recovery, and their overall experience with the system.

\begin{figure*}[t!]
    \centering
    \begin{subfigure}{0.2\textwidth}
        \centering
        \includegraphics[clip, trim= {0cm, 10cm, 1.5cm, 10cm}, width=\textwidth]{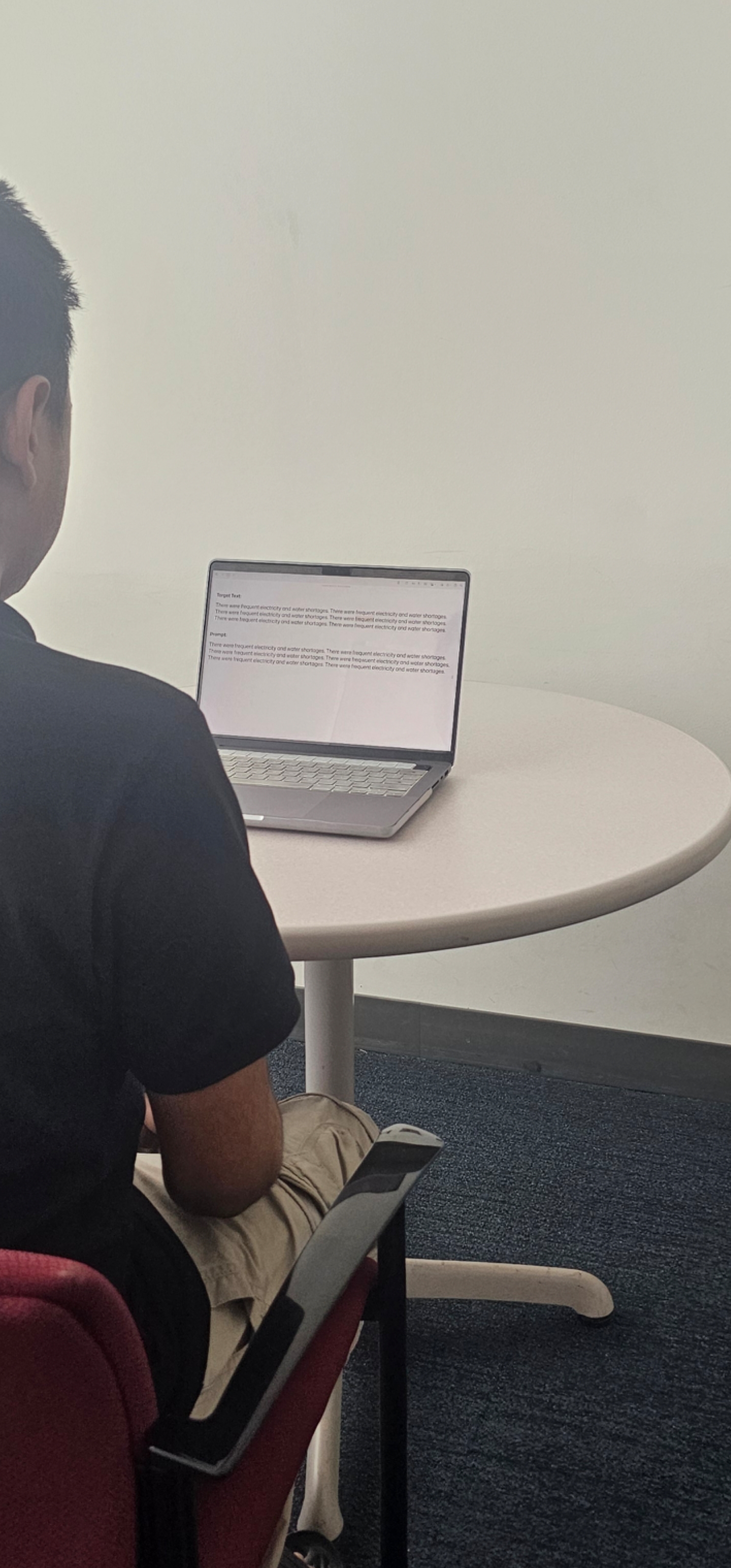}
        \caption{}
        \label{fig:setup-overview}
    \end{subfigure}
    \hfill
    \begin{subfigure}{.79\textwidth}
        \centering
        \includegraphics[clip, trim= {0cm, 14cm, 0cm, 1.5cm}, width=\textwidth]{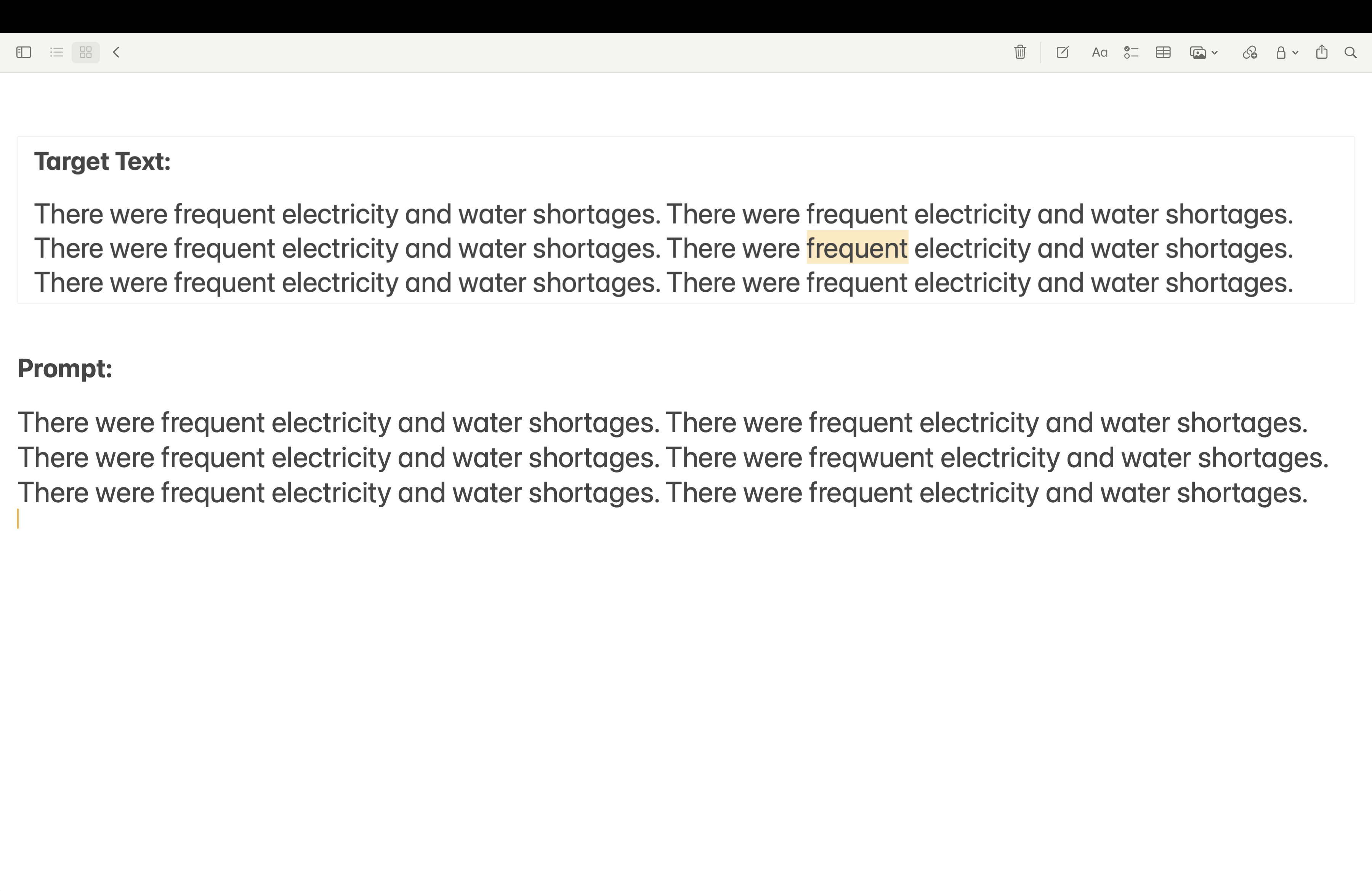}
        \caption{}
        \label{fig:task-overview}
    \end{subfigure}
	\vspace{-10pt}
    \caption{Study setup showing (a) the physical arrangement with the participant seated before the laptop running Voice Control and (b) the interface displaying a sample correction task with target text and prompt.}
    \Description{Study setup showing (a) the physical arrangement with the participant seated before the laptop running Voice Control and (b) the interface displaying a sample correction task with target text and prompt.}
    \label{fig:study-setup}
    \vspace{-10pt}
\end{figure*}

\subsection{Study Setup and Apparatus}
We conducted the study in a quiet room with participants seated before a MacBook Pro running Apple Voice Control (Figure~\ref{fig:setup-overview}) as the VUI. For presenting correction tasks, we used the Notes application, a popular text editing environment for dictation users. As shown in Figure~\ref{fig:task-overview}, each trial displayed both the \texttt{\small Target Text} (correct version with the target word highlighted) and the \texttt{\small Prompt} (erroneous version to be corrected). Each prompt contained exactly one error compared to its target.

To create a realistic correction scenario where errors might appear anywhere within a body of text, we followed Zhao et al.'s~\cite{zhao2022eyesay} approach by embedding each erroneous sentence within a paragraph of 1-7 sentences. These additional sentences—identical copies of the correct version—served as distractors during target acquisition. For example, in Figure~\ref{fig:task-overview}, the highlighted word \textbf{"frequent"} in the target appears as \textbf{"freqwuent"} in the prompt, surrounded by correctly spelled instances of the same sentence.

We deliberately avoided suggesting specific commands for any task, instead observing which instructions participants naturally selected to accomplish each correction.

\subsection{Study Procedure}
After obtaining informed consent and demographic information, we began with a training session on the VUI system and its supported text correction commands. Participants practiced with these commands until they reported feeling comfortable with the system. We then provided sample trials for each task type to ensure familiarity with the correction process.

For the main study, we presented 20 trials in randomized order to each participant. The randomization was performed at two levels: (i) task assignment was randomized such that participants received correction tasks (insertion, deletion, replacement, and typo correction) in varying sequences rather than grouped by type, and (ii) within each task type, the specific trial instances were randomized. This ensured that no two participants experienced the same task sequence and that learning effects were not confounded with specific task types.

In each trial, participants first reviewed the \texttt{\small Target Text} and \texttt{\small Prompt} to understand what and where to correct. Once they acknowledged understanding the required correction, we switched to another Notes tab containing only the \texttt{\small Prompt} text. Participants could revisit the target if needed, and could use as many commands as necessary until the text matched the target. We advanced to the next trial only when participants confirmed completion of the current one, with rest periods available between trials. Following the correction tasks, we conducted a semi-structured interview to explore participants' specific strategies and experiences.

\subsection{Data Analysis}
We transcribed and analyzed the data through iterative coding~\cite{semi-structured-interview}, with all authors participating in weekly research meetings to refine the codebook, identify key concepts, organize categories, and resolve disagreements. Our initial coding cycle identified low-level descriptive codes representing specific command and task strategies.
In subsequent cycles, we merged conceptually similar codes into broader categories that revealed primary themes. This refinement process ultimately yielded four key themes that we describe next. 

%% file: source/sections/4_pilot_study_findings.tex
\subsection{Study Findings}
Our formative study revealed critical usability challenges with fixed-format voice commands that inform the design of \sysname{}. We organize these findings around four themes: target acquisition, instruction planning and execution, task completion strategies, and overall system usability.

\subsubsection{Target Acquisition Challenges}

\label{sec:target-acquire}
We found that target acquisition with voice commands was cumbersome and inconvenient in two cases: when the target had multiple occurrences or contained a typo/out-of-vocabulary words. 
If \partis{} used a \texttt{\small select <phrase>} command where the \texttt{\small <phrase>} occurred multiple times, the system considered all occurrences as the potential target and tagged each of them with a numeric index starting from 1. Then \partis{} issued additional \texttt{\small choose <number>} or just the \texttt{\small <number>} command to disambiguate the particular \texttt{\small <phrase>} they wanted to select. 
All \partis{} reported this process to be time-consuming.
P15 thought it was extra decision-making for him as he had to look for the exact instance out of the duplicates he wanted to select.
P11 liked numbers for disambiguation but also wanted to have the numbers during selection, not only during disambiguation.

When the target was a typo or an out-of-vocabulary word, \partis{} had difficulty acquiring them as the ASR did not recognize them. The first strategy that all \partis{} tried was to utter the typo as a word, if possible, by dividing it into syllables. For example, they uttered the typo `cintrol' as a two-syllable word `cin'--`trol' and used that as the \cmdarg{}, but the ASR could not recognize it. The most common and successful strategy \partis{} employed to select a typo required three commands in most cases:  
(i) use a \selectcmd{} command to select a word adjacent to the typo; 
(ii) use a \choosecmd{} command to disambiguate (if duplicate);
and (iii) another \selectcmd{} command with \texttt{\small previous/next word} as \ctx{} to select the typo.
Another option was to spell out the typo using the \texttt{\small spelling mode}.
However, spelling longer words containing typos (e.g., `freqwuent' containing 9 characters)
was prone to misrecognition because if the system could not recognize a single character, the selection failed.
Therefore, \partis{} needed multiple attempts with longer typos.

For target acquisition, \partis{} preferred using the \selectcmd{} command predominantly over the \texttt{\small move} command. In addition, they also preferred easier \ctx{} over larger ones, such as \texttt{\small select previous word} over \texttt{\small select previous <number> words}. We never encountered commands such as \texttt{\small move left <number> words}, which required a distance calculation. The only \texttt{\small move} commands \partis{} used were \texttt{\small move before/after <phrase>}. In other words, selection-based target acquisition was more convenient than navigation-based acquisition, which is also corroborated in prior work~\cite{sears2003hands}.

\subsubsection{Command Planning and Execution Barriers}

\begin{table*}[t!]
    \centering
    \renewcommand{\arraystretch}{1.6}
    \begin{tabular}{l|p{6cm}|p{3.5cm}|p{3.5cm}}
         \textbf{Category} & \textbf{Description} & \textbf{Incorrect Command} & \textbf{Correct Command} \\\toprule
         \multirow{2}{2cm}{Swap \cmd{}} & \multirow{2}{6cm}{Swapping one valid \cmd{} for another valid one, particularly common for select and choose.} & \texttt{\small select <number>} & \texttt{\small choose <number>} \\\cline{3-4}
             &  & \texttt{\small choose <phrase>} & \texttt{\small select <phrase>}\\\midrule
         \multirow{4}{*}{Substitute \cmd{}} & \multirow{4}{6cm}{Substituting one valid \cmd{} with another invalid synonym.} & \texttt{\small add <phrase> before <phrase>} & \texttt{\small insert <phrase> before <phrase>} \\\cline{3-4}
             &  & \texttt{\small fix <phrase>} & \texttt{\small correct <phrase>}\\\cline{3-4}
            &  & \texttt{\small remove <phrase>} & \texttt{\small delete <phrase>}\\\midrule
        \multirow{5}{*}{Substitute \ctx{}} & \multirow{5}{6cm}{Substituting one valid \ctx{} with another invalid synonym.} & \texttt{\small select left word/select word before} & \texttt{\small select previous word} \\\cline{3-4}
             &  & \texttt{\small select right word/select word after} & \texttt{\small select next word}\\\cline{3-4}
            &  & \texttt{\small replace <phrase> to/using <phrase>} & \texttt{\small replace <phrase> with <phrase>}\\\midrule
        \multirow{3}{*}{Substitute template} & \multirow{3}{6cm}{Substituting one valid template with another template with more or fewer components.} & \texttt{\small insert <phrase>} & \texttt{\small insert <phrase> before <phrase>} \\\cline{3-4}
            &  & \texttt{\small delete <phrase> before <phrase>} & \texttt{\small delete <phrase>}\\\midrule
        \multirow{2}{*}{Ignore deictic args} & \multirow{2}{6cm}{Ignoring deictic arguments such as this/that after selection or for commands that do not have any particular reference.} & \texttt{\small delete} & \texttt{\small delete that} \\\cline{3-4}
            &  & \texttt{\small undo/redo} & \texttt{\small undo that/redo that}\\\midrule
        \multirow{3}{*}{Add deictic args} & \multirow{3}{6cm}{Adding deictic arguments such as this/that after selection in four-component commands.} & \texttt{\small insert <phrase> before that} & \texttt{\small insert <phrase> before <phrase>} \\\cline{3-4}
            &  & \texttt{\small replace that with <phrase>} & \texttt{\small replace <phrase> with <phrase>}\\\midrule
        \multirow{3}{*}{Missing args} & \multirow{3}{6cm}{Missing arguments in four-component commands after selection to reduce command complexity.} & \texttt{\small insert <phrase> before} & \texttt{\small insert <phrase> before <phrase>} \\\cline{3-4}
            &  & \texttt{\small replace with <phrase>} & \texttt{\small replace <phrase> with <phrase>}\\\midrule
        \multirow{2}{*}{Natural utterance} & \multirow{2}{6cm}{Uttering fixed-format commands naturally, creating minor variations.} & \texttt{\small choose number <number>} & \texttt{\small choose <number>} \\\cline{3-4}
            &  & \texttt{\small correct the selected word} & \texttt{\small correct that}\\
        \bottomrule
    \end{tabular}
    \caption{Categories of incorrect commands identified during the study and their correct versions. Note that all uttered commands are minor variations of the fixed-format command templates, which are not supported by VUIs.}
    \Description{Categories of incorrect commands identified during the study and their correct versions. Note that all uttered commands are minor variations of the fixed-format command templates, which are not supported by VUIs.}
    \label{tab:incorrect-commands}
\end{table*}

Successfully executing voice commands required participants to select a valid combination of command components, retrieve each component's value from short-term memory, and complete the utterance before timeout. 

Initial attempts frequently failed because \partis{} did not plan their utterances to match the system's expectations. With four possible components creating 16 potential combinations --- only 6 of which are valid (Figure~\ref{fig:cmnd-comb}) --- \partis{} faced a 40\% chance of selecting a valid combination. This probability decreased further as different commands expected specific component combinations.

Participants described commands as ``restrictive,'' ``programming-like,'' and ``unnatural.'' They struggled with several structural issues, including component quantity mismatches, synonymous commands, context word interchanging, and natural speech patterns. As P3 observed: \textit{"It is not natural; it is not the way we speak. It is like we need to change ourselves. It is like you have to follow very specific instructions."}

Even when participants correctly structured their commands, timeouts frequently disrupted execution. The process of retrieving component values from different sources while maintaining the precise utterance pace proved challenging. This difficulty intensified for complex commands requiring context words that established relationships between arguments.

P2 articulated this challenge: \textit{``I have to say the exact command... At the same time, I have to also say it fast... I have to process my sentence to make up the command, and that too in a precise time. That's a bit tricky!''}

First attempts typically failed due to attention division between component retrieval and coherent utterance. Most participants initially required three attempts per command (spiral depth of three), eventually improving to two attempts with practice. Unsurprisingly, timeout frequency increased with command complexity, leading most participants to prefer simpler two-component commands.

We analyzed the uttered commands from our study that did not work and categorized them into 8 categories. Table~\ref{tab:incorrect-commands} shows these categories, the incorrect commands, and their correct versions. Notice that the incorrect commands are minor variations of the correct ones that \partis{} thought made sense and would give flexibility and ease-of-use if supported by the system.

\subsubsection{Task Completion Strategies}
\begin{figure*}[t!]
    \centering
    \begin{subfigure}{\linewidth}
        \centering
        \includegraphics[clip, trim= {2.2cm, 3cm, 3.4cm, 2.9cm}, width=\linewidth]{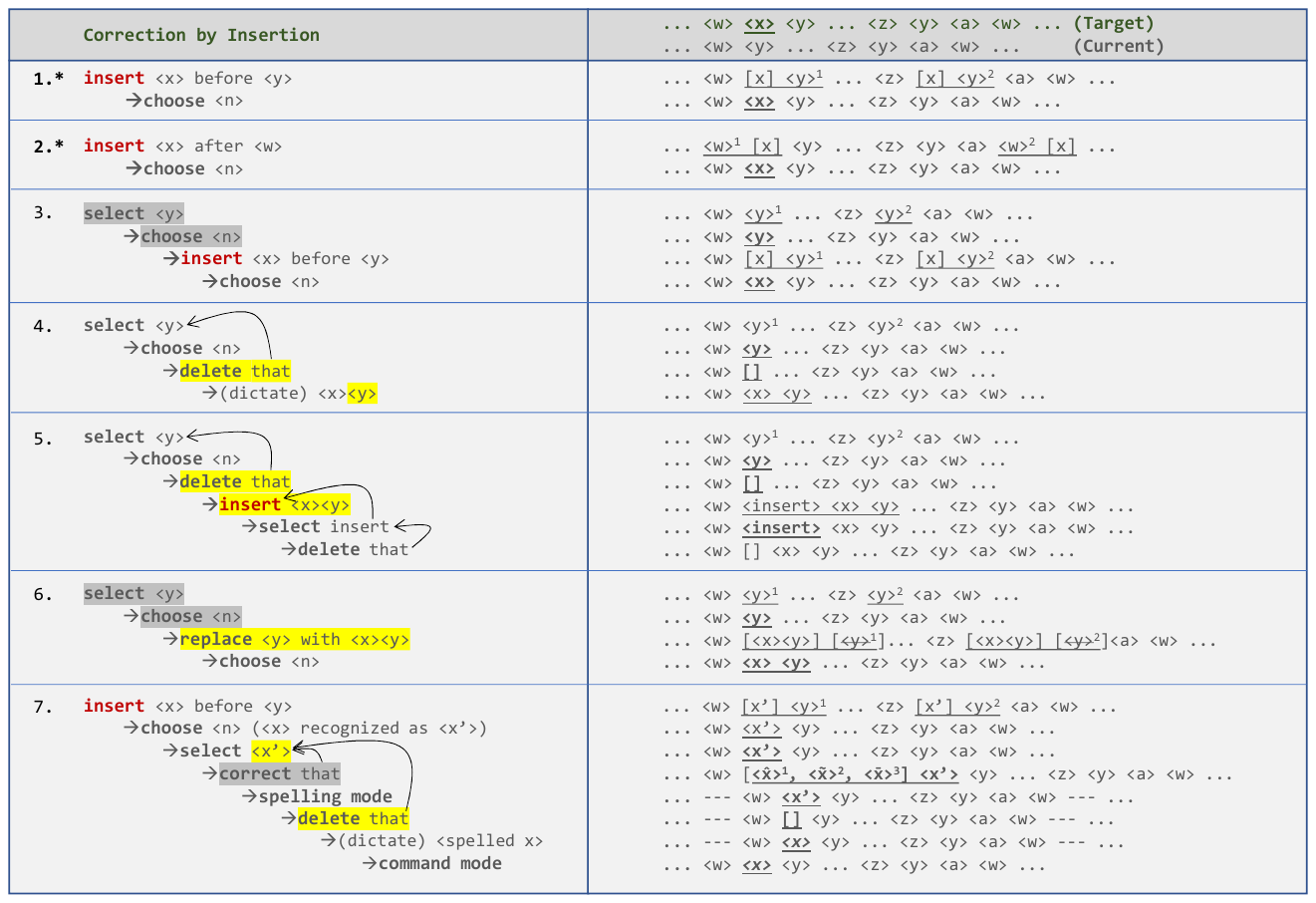}
    \end{subfigure}
	\vspace{-12pt}
 \caption{Common instances of command sequences used by \partis{} to accomplish corrections by insertions on the left and the workflow of the system on the right. An asterisk (*) indicates the optimal command sequences. The target text and the current state are shown at the top right. An underline is used to indicate where editing occurred. A square brace ([]) indicates a temporary buffer the system uses internally (e.g., all potential insertions until disambiguated).}
 \Description{Common instances of command sequences used by participants to accomplish corrections by insertions on the left and the workflow of the system on the right. An asterisk (*) indicates the optimal command sequences. The target text and the current state are shown at the top right. An underline is used to indicate where editing occurred. A square brace ([]) indicates a temporary buffer the system uses internally (e.g., all potential insertions until disambiguated).
 }
	\label{fig:insert-trace}
\vspace{-12pt}
\end{figure*}

The open-ended nature of voice commands required participants to strategize their approach before execution. We observed several consistent patterns:

Participants instinctively began with target acquisition across all tasks, mirroring desktop interaction patterns where selection precedes action. However, selected targets did not persist between commands, creating redundant work. For example, when inserting text near a selected and disambiguated location, the system would require re-disambiguation after the insert command, rendering the initial selection meaningless.

Over time, participants developed a consistent two-step strategy of acquisition followed by correction, both using two-component instructions. For example, they would select a word and then use a deictic reference like \texttt{\small delete that} to modify it. When commands didn't support such references (as with \insertcmd{} and \replacecmd{}), participants resorted to workarounds like deleting targets and re-dictating replacements.

When facing errors, participants employed various recovery strategies: using \texttt{\small undo that}, switching to \texttt{\small command mode}, or using \texttt{\small spelling mode} for unrecognized words. Most notably, once participants found a successful approach --- even if suboptimal --- they adhered to it, following an ``if it works, don't break it'' strategy to minimize errors.

We analyzed how \partis{} completed the four correction tasks in-depth and identified how they performed the same correction tasks in different ways. We outlined the seven different ways \partis{} completed the insertion task in Figure~\ref{fig:insert-trace} along with a visual representation of how each command modifies the text through Voice Control. Notice the persistence of selection at the beginning that does not benefit due to additional disambiguation. Additionally, the insertion was often done through other commands that \partis{} kept using if successful.

\subsubsection{System Usability Issues}
Beyond specific command challenges, participants identified several broader usability concerns. They reported having to proactively control the microphone to prevent unintended dictation, particularly when thinking aloud or reading text to themselves. This often led to cascading errors when expressions of frustration themselves became dictated text.

Participants expressed mixed preferences regarding correction modes, with most seeking to minimize mode switching. While half the participants valued both dictation and command modes equally, others preferred one over the other based on flexibility and error prevention needs. The \texttt{\small spelling mode} was generally considered a last resort despite its utility for specific tasks.

All participants highlighted insufficient system feedback as a major issue. They often couldn't determine whether an uttered command had been recognized, whether a timeout had occurred, or which mode was currently active. As P7 suggested, the system should ``clearly distinguish between just listening and listening to a command'' through visual cues like live transcription of recognized command components.

\subsection{Discussion: Designing Better Voice-Based Text Correction}
Our findings reveal a fundamental mismatch between legacy VUI command structures and users' cognitive models of speech interaction. Unlike programming, which is primarily text-based, speech is naturally more verbose, erroneous, and disorganized. The rigid instruction formats fail to accommodate these characteristics, creating a substantial cognitive burden.

Based on our study, we propose six design guidelines for improving speech-based text correction:

\begin{enumerate}
    \item \textbf{Simplify target acquisition} by extending numeric indexing to all words in text, allowing direct selection through \texttt{\small select <number>} commands.
    
    \item \textbf{Prioritize two-component instructions} to reduce cognitive load, with systems inferring contextual relationships (like insertion position) based on text analysis or user preferences.
    
    \item \textbf{Maintain selection context} across commands, preserving target information from previous operations to eliminate redundant selections.
    
    \item \textbf{Minimize timeouts} by extending thresholds and implementing user-defined completion phrases like "over" or "end" to signal command completion.
    
    \item \textbf{Support minor variations} in commands through more flexible interpretation, potentially leveraging large language models (LLMs) to understand intent despite syntactic variations.
    
    \item \textbf{Enhance system feedback} through clear visual and auditory cues indicating recognition status, active mode, and detected command components.
\end{enumerate}

These guidelines inform our development of \sysname{}, a shimming layer that addresses these limitations while preserving compatibility with existing legacy VUI systems.

%% file: source/sections/5_system_design.tex
\section{Designing \sysname{}: An Adaptive Shimming Layer}

Our analysis of voice commands indicates that legacy VUIs provide a comprehensive set of commands for interaction and text correction. However, our formative study revealed that users face difficulty using these comprehensive command sets due to fixed formats, timeouts, and insufficient feedback. These systems are black-box and difficult to modify, while creating a completely new VUI with all the functionality that existing VUIs support is costly and cumbersome. Therefore, in this work, we aimed to leverage the capabilities of existing VUIs while also allowing users to speak command variations we identified during our formative study without modifying them.

This led us to design \sysname{}, an adaptive shimming layer that mediates between users and legacy VUI systems.
The concept of shimming layers originated in software engineering as the Adapter pattern~\cite{gamma1995design}. \sysname{} has the following design goals:

\begin{itemize}
    \item \textbf{Interfacing with Existing Black-box VUIs}: Existing VUIs, such as Voice Control, are widely available and offer comprehensive features. We aim to interface with these systems, treating them as black boxes, allowing us to leverage their features.
    \item \textbf{Supporting Command Variations}: We aim to support command variations, allowing users to speak commands naturally and align them with fixed-format commands using large language models.
    \item \textbf{Enhancing User Experience}: Our goal is to enhance user experience by providing additional feedback to keep users informed while using voice commands and giving them sufficient time for planning and uttering, as we identified during our formative study.
\end{itemize}

\subsection{System Architecture and Workflow}
\sysname{} has two primary components: a command adapter and a command whisperer (see Figure~\ref{fig:teaser}).
Currently, it runs as a web application in a browser. 

\subsubsection{Command Adapter}
The command adapter intercepts voice input before it reaches the legacy VUI system. This component serves three functions:

\begin{itemize}
    \item \textbf{Command recognition}: It captures and transcribes user utterances in real-time using Web Speech API\footnote{\url{https://developer.mozilla.org/en-US/docs/Web/API/Web\_Speech\_API/Using\_the\_Web\_Speech\_API}} and provides immediate visual feedback on detected speech.
    
    \item \textbf{Timeout management}: It implements a customizable inter-/intra-component timer (initially set to 3 seconds), giving users sufficient time to plan and articulate commands without pressure.
    
    \item \textbf{Command transformation}: Using an LLM (Claude\-3.5-Sonnet API, with temperature=0), it maps users' natural language commands to syntactically valid formats required by the legacy system. 
    For each command, the LLM receives: the user's transcribed utterance; the currently selected text (if available); and a history of the five most recently executed commands. It then extracts necessary components (\cmd{}, \cmdarg{}, \ctx{}, and \ctxarg{}) and reformulates them into VUI-specific syntax.
\end{itemize}

The system maintains contextual awareness by tracking selection state through command history. When users issue a \texttt{\small select <phrase>} command, \sysname{} stores it in its internal cache. For commands like \texttt{\small choose <n>}, it references command history to determine whether to preserve the current selection, mimicking Voice Control's default behavior.

\begin{figure*}[t!]
    \centering
    \includegraphics[clip, trim= {0cm, 4.2cm, 0cm, 4cm}, width=\linewidth]{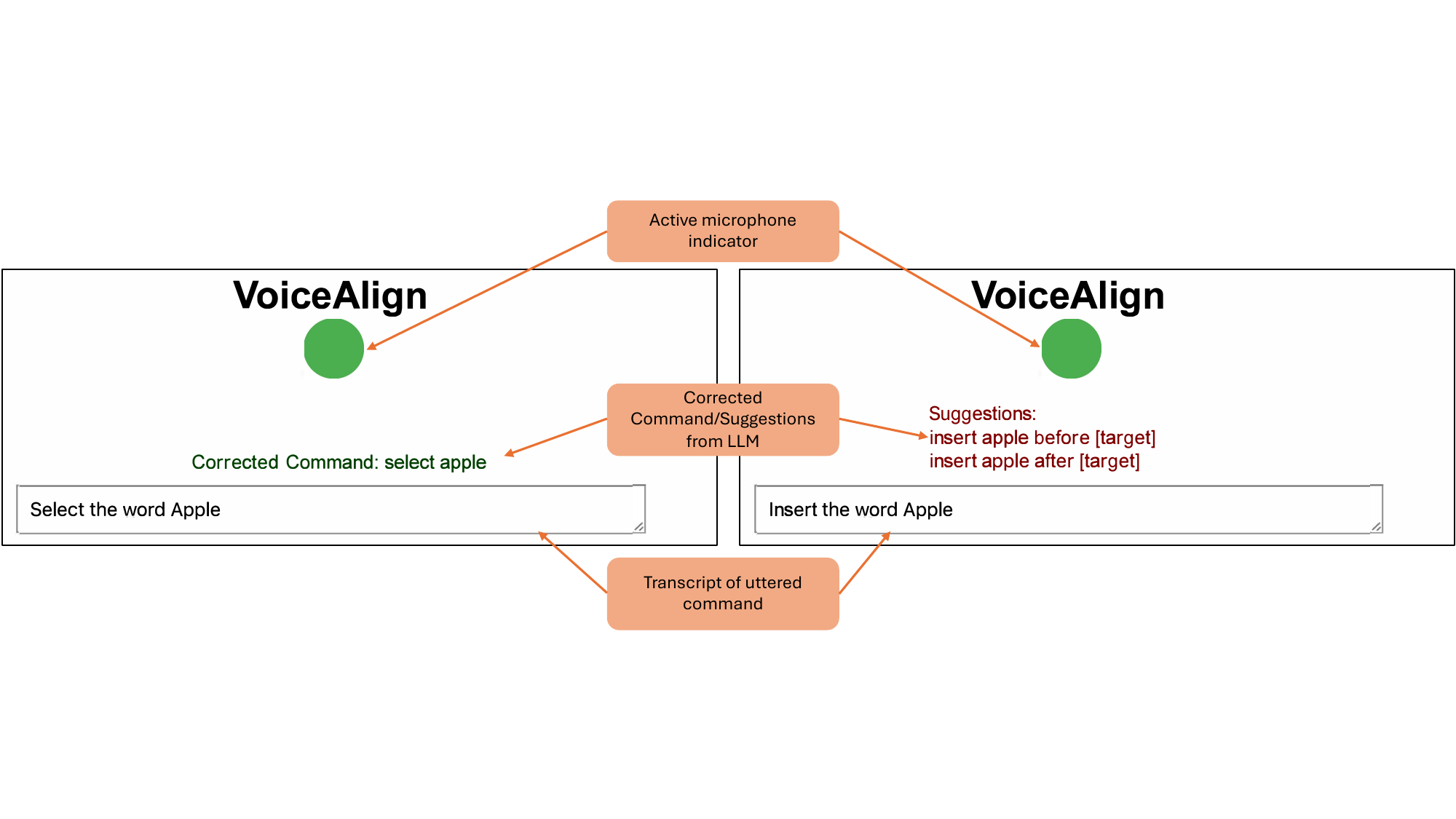}
    \label{fig:llm-correct}
	\vspace{-12pt}
 \caption{\sysname{} interface containing an indicator when the microphone is active, an input box to display the transcript of users' uttered commands in real-time, and the output from the LLM providing the correct command or a list of suggestions. (Left) An example of a command correction, where \texttt{`Select the word Apple'} is transformed to the syntactically valid \texttt{`select apple'} by removing extraneous words while preserving core components. (Right) An example of a command suggestion, where \texttt{`Insert the word Apple'} lacks required context parameters, prompting the system to offer structured guidance on how to complete the command.}
 \Description{VoiceAlign interface containing an indicator when the microphone is active, an input box to display the transcript of users' uttered commands in real-time, and the output from the LLM providing the correct command or a list of suggestions. (Left) An example of a command correction, where 'Select the word Apple' is transformed to the syntactically valid select apple by removing extraneous words while preserving core components. (Right) An example of a command suggestion, where 'Insert the word Apple' lacks required context parameters, prompting the system to offer structured guidance on how to complete the command.}
 \label{fig:interface}
\end{figure*}

When a user speaks, the adapter displays an active microphone icon indicating active listening, shows real-time transcription, and processes the input once the timer expires (see Figure~\ref{fig:interface}).

\subsubsection{Command Whisperer}
The command whisperer creates a virtual audio channel connecting the adapter to the legacy VUI through two components:

\begin{itemize}
    \item \textbf{Virtual channel creation}: It establishes a loopback audio pathway using BlackHole\footnote{\url{https://github.com/ExistentialAudio/BlackHole}} that redirects system audio output to serve as input for the legacy VUI.
    
    \item \textbf{Text-to-speech relay}: Once the adapter produces a corrected command, the whisperer converts it to speech using the system TTS engine and plays it through the virtual channel to the legacy VUI.
\end{itemize}

This approach allows \sysname{} to function as a transparent layer, with the legacy VUI unaware that commands are coming from an intermediary rather than directly from the user.

\subsection{LLM Prompting Strategy}
Our implementation uses a carefully crafted prompting strategy to ensure reliable command transformation while avoiding hallucinations or fabricated commands that could disrupt the user experience.
We developed the prompt iteratively based on users' utterances and logs from the formative study, particularly those that resulted in errors (\S\ref{sec:pilot}), as well as our generalized command template developed during the analysis of legacy VUIs (\S\ref{sec:analysis}). 

\subsubsection{Role Definition and Command Analysis}
We designed a role-based prompt where the LLM functions as an "advanced command corrector" specialized in text editing operations. The prompt instructs the model to:

\begin{enumerate}
    \item Analyze the command to identify its intended operation type.
    \item Determine the required components for that operation.
    \item Extract arguments directly from the user's utterance without inference.
    \item Reconstruct a syntactically valid command only when all required components can be reliably extracted.
\end{enumerate}

To prevent command fabrication, we explicitly constrain the model to use only consecutive words directly from the user's utterance for arguments, prohibiting the generation of new content except in specific, well-defined cases.

\subsubsection{Command Completion and Error Handling}
For commands that reference previous selections (e.g., \texttt{\small insert <phrase> before that}), we provided explicit guidelines for resolving deictic references based on selection history. This enables \sysname{} to support natural command sequences while maintaining syntactic precision.

When a command lacks sufficient information for reliable transformation, the system shifts to suggestion mode rather than attempting correction. As shown in Figure~\ref{fig:interface} (right), it provides structured guidance on how to complete the command, maintaining user trust by avoiding potentially incorrect transformations.

\subsubsection{Confidence Assessment}
To further enhance reliability, our prompt requires the model to:

\begin{itemize}
    \item Show explicit reasoning for each transformation step.
    \item Assign a confidence score (0-100) to each correction attempt.
    \item Only apply corrections with high confidence scores.
\end{itemize}

This approach, inspired by Li et al.'s work on improving LLM precision~\cite{li2024think}, ensures that \sysname{} prioritizes accuracy over coverage, preserving user trust in the system's transformations.

%% file: source/sections/6_findings.tex
\section{Evaluation of \sysname{}}
We conducted a summative study to evaluate the effectiveness and user experience of \sysname{} when working as a shim layer between users and VUIs. This section describes the study setup, findings, and comparisons with the earlier study.

\subsection{Participants, Study Design, and Setup}
We recruited 12 \partis{} (10 males, 2 females, ages 26-32) using the same recruitment criteria as our formative study. Participants, anonymized as P1 through P12, performed identical text correction tasks using \sysname{}: 20 erroneous sentences across four task types. The study setup mirrored our formative evaluation with one key difference: participants could now see the \sysname{} interface displayed above the Notes application, providing supplementary feedback alongside Voice Control's native responses.

\subsection{Study Procedure and Data Analysis}
After collecting consent and demographic information, we introduced participants to the \sysname{} interface and provided them with Voice Control commands. 

Participants completed 20 trials of the four correction tasks in random order. Again, the randomization was performed at both the task level (trials for the tasks were randomized) and within each task type (the specific trials were randomized) to mitigate the learning effect. 

Rather than instructing participants to deliberately vary their command structures, we observed their natural interactions with Voice Control, noting incorrect commands, \sysname{}'s responses, and participants' subsequent behaviors.

Following the correction tasks, we conducted semi-structured interviews and administered the NASA-TLX questionnaire to measure task load. Each session lasted 60-75 minutes, was video-recorded for analysis, and concluded with participants receiving a \$20 Amazon gift card as compensation. Our data analysis followed the same approach used in the formative study.

\subsection{Quantitative Results}
\begin{figure}[t!]
    \centering
    \begin{subfigure}{0.49\linewidth}
        \centering
        \includegraphics[width=\linewidth]{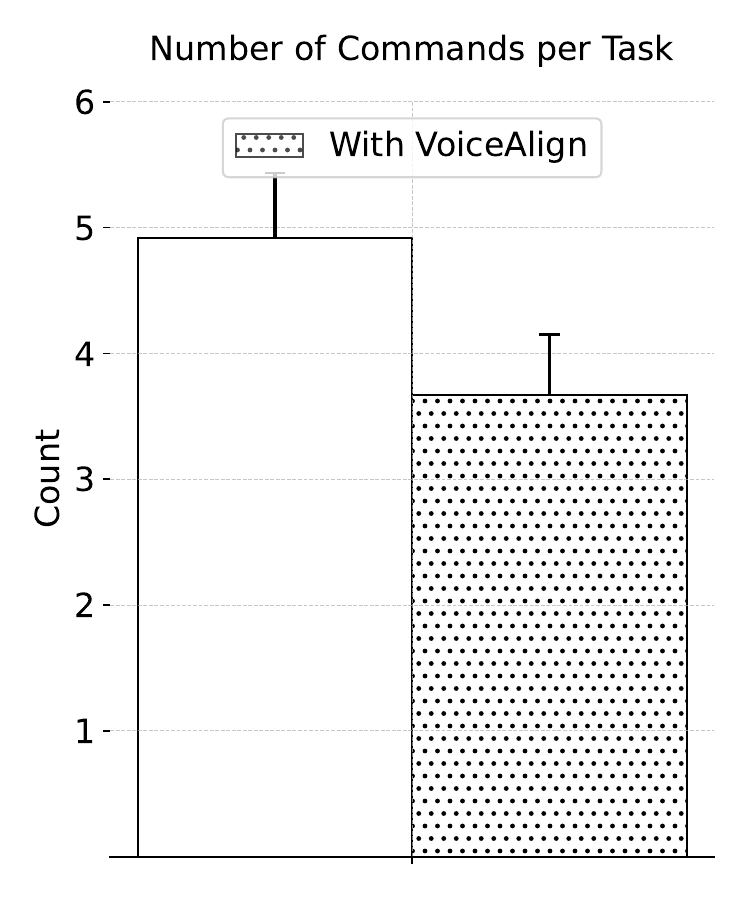}
        \caption{}
        \label{fig:num-commands}
    \end{subfigure}
    \begin{subfigure}{.49\linewidth}
        \centering
        \includegraphics[width=\linewidth]{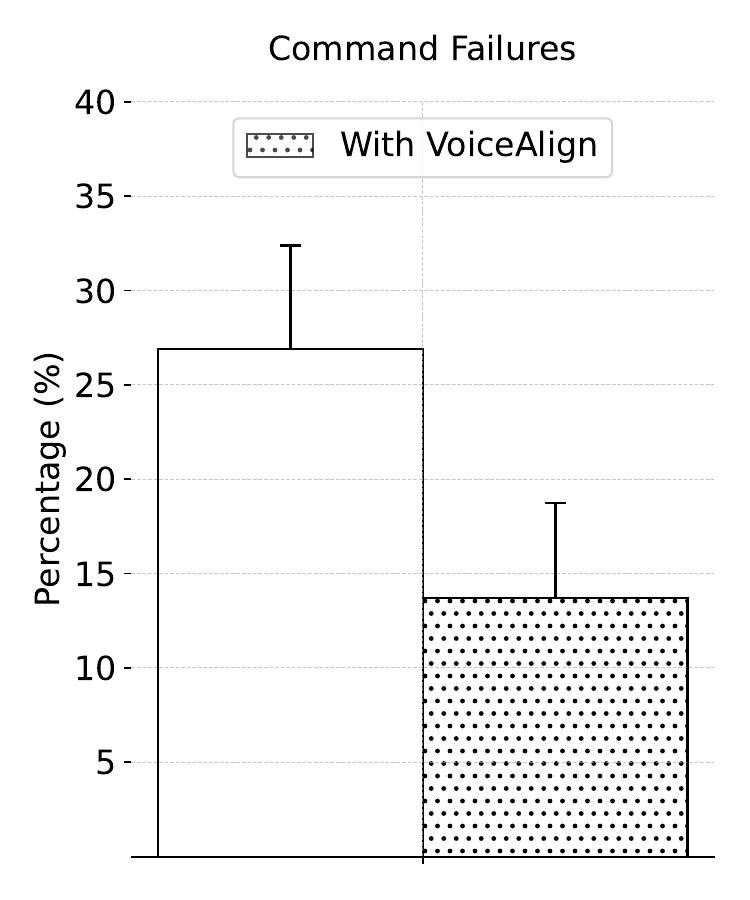}
        \caption{}
        \label{fig:failed_commands}
    \end{subfigure}	
    \caption{(a) Average number of commands \partis{} used to perform each text correction task. (b) Percentage of commands that failed out of all commands by \partis{}.}
    \Description{(a) Average number of commands participants used to perform each text correction task. (b) Percentage of commands that failed out of all commands by participants.}
    \label{fig:quant}    
\end{figure}

\subsubsection{Number of Commands per Task}
As shown in Figure~\ref{fig:num-commands}, \sysname{} reduced the average number of commands needed per correction task by 25\% --- from 4.92 (SD: 0.51) with Voice Control to 3.67 (SD: 0.48) with \sysname{} + Voice Control. An independent samples t-test confirmed this difference was statistically significant ($t(26) = 6.56, p <.001$). By preventing command failures, \sysname{} enabled participants to complete tasks with ease.

\subsubsection{Command Failures}
Figure~\ref{fig:failed_commands} illustrates the percentage of failed commands across systems. With Voice Control alone, 26.90\% (SD: 5.48\%) of commands failed, compared to 13.70\% (SD: 5.05\%) when \sysname{} adapted commands before passing them on to Voice Control. An independent samples t-test confirmed this difference was statistically significant ($t(26) = 6.526, p <.001$).

We further analyzed the command failures for Voice Control with and without \sysname{}. We categorized the failures from Voice Control without \sysname{} into three categories: timeouts, incorrect command syntax, and recognition errors. Notably, timeouts caused 33\% of command failures in Voice Control. 35.8\% of the failures were due to incorrect command syntax, shown in Table~\ref{tab:incorrect-commands}, and the rest of the commands failed due to misrecognition.

For \sysname{} + Voice Control, the majority of the command failures (76.5\%) were due to incorrect recognition. Approximately 20\% of the failures occurred due to incorrect command syntax, which \sysname{} could not correct due to missing arguments. However, in such cases, users were informed as \sysname{} provided suggestions to the users, which they could use in the next attempt to correct their command. No timeouts occurred with \sysname{} due to our adjusted timeout mechanism in the shimming layer. \sysname{} could commonly correct commands with swapped or substituted \cmd{} or \ctx{}, natural utterances, and ignored/added deictic args (Table~\ref{tab:incorrect-commands}). However, we noticed a few occurrences where \sysname{} provided a suggestion for correctly uttered commands containing `that' as an argument (e.g., \texttt{\small correct that}) without prior selection, where the model suggested selecting first.

\subsection{Subjective Findings}
\begin{figure}[t!]
    \centering
    \begin{subfigure}{\linewidth}
        \centering
        \includegraphics[width=\textwidth]{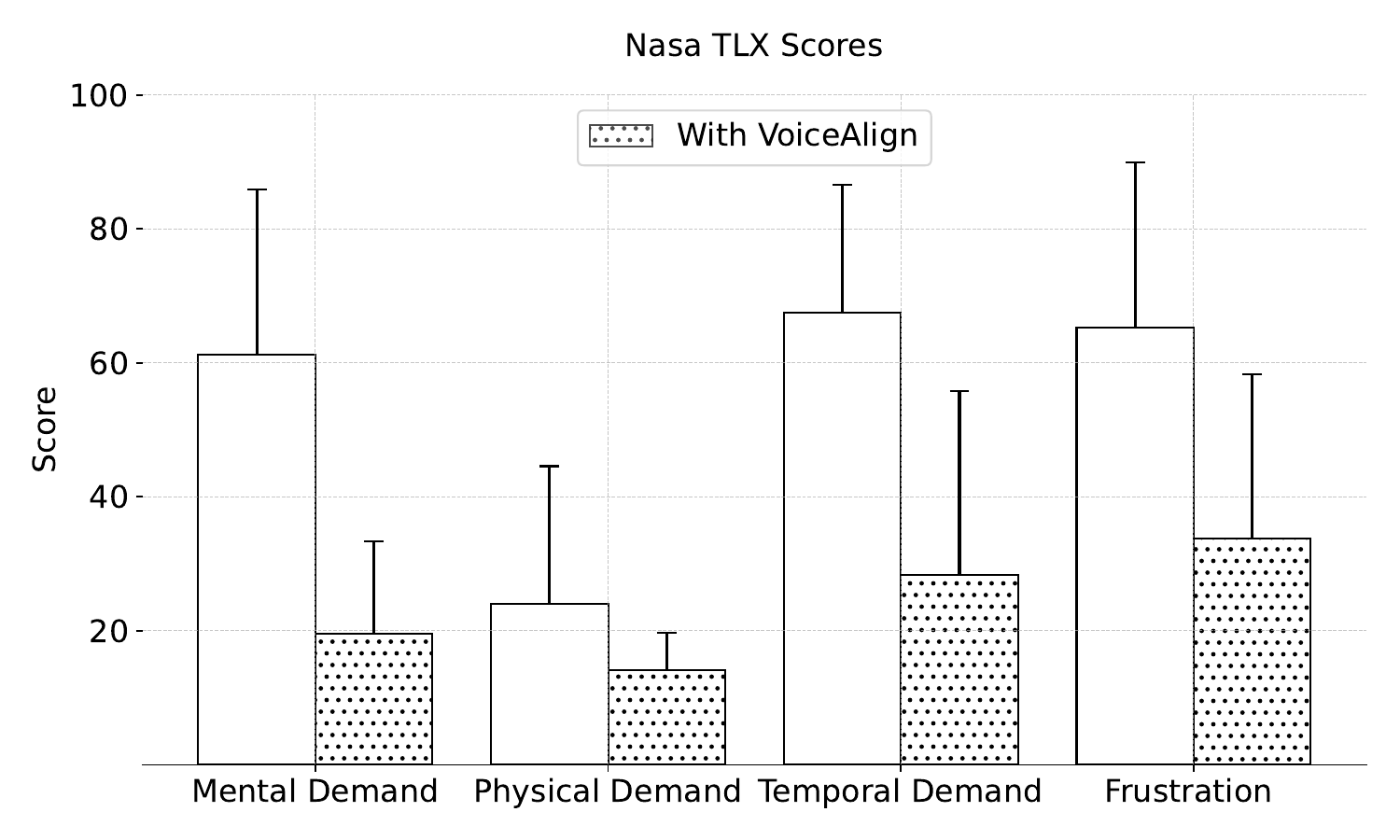}
    \end{subfigure}
	\vspace{-10pt}
    \caption{Comparison of NASA-TLX Scores with and without \sysname{}.}
    \Description{Comparison of NASA-TLX Scores with and without VoiceAlign.}
    \label{fig:nasa-tlx}
    \vspace{-10pt}
\end{figure}

\subsubsection{NASA TLX Score}
The NASA-TLX scores (Figure~\ref{fig:nasa-tlx}) revealed that \sysname{} reduced mental demand, physical demand, temporal demand, and frustration when paired with Voice Control. A Mann-Whitney U test confirmed significant reductions in mental demand ($Z = -3.692, p < .001$), temporal demand ($Z = -3.180, p = .001$), and frustration ($Z = -2.687, p = .007$), though the reduction in physical demand was not statistically significant ($Z = -1.448, p = .148$).

\subsubsection{\sysname{} Adapted Voice Commands Successfully}
All participants responded positively to \sysname{}'s automatic command correction. When P1 accidentally issued \texttt{\small select <n>} instead of \texttt{\small choose <n>} and saw the command automatically corrected, he exclaimed: \textit{``Awesome! That's how it should be!''}. We observed similar enthusiasm across participants when they realized they could deviate from exact command formats while still achieving their goals.

Some participants adapted their command strategies based on \sysname{}'s capabilities. For example, P5 discovered that after selecting a word, she could refer to it in subsequent commands using deictic references (e.g., \texttt{\small replace that with <phrase>}). She commented:

\begin{quote}
\emph{``This makes more sense to me. I can use the same strategy for other commands. I can select a word and then say delete that. Then why can't I say `replace that with [<phrase>]'? It helps me because I do not have to think about two words for one command. I would rather replace that than say two words.''}
\end{quote}

This finding aligned with our formative study, where participants reported that combining multiple arguments in a single command increased mental load. \sysname{} also successfully handled synonymous commands (e.g., \texttt{\small fix that} for \texttt{\small correct that}) and removed verbal fillers and false starts (e.g., ``Okay...\texttt{\small select <phrase>}'').

We observed a few edge cases where \sysname{} incorrectly discarded valid commands because the LLM lacked sufficient context, particularly with deictic references following selection commands that modified context in ways the LLM couldn't track. 

\subsubsection{\sysname{} Mitigated Timeout Errors}
All participants successfully issued commands within our extended 3-second timer, eliminating the timeout errors. None reported difficulties planning or articulating command components, which corresponded with significantly decreased mental and temporal demand scores.

Two participants (P4 and P12) suggested that the timer could adaptively adjust based on user expertise, allowing longer intervals for novices and shorter ones for experienced users. P4 noted that for multi-component commands, the system could potentially use syntactic cues to recognize completion: \textit{``If I already said more than four words, chances are I am done with my command, so the system does not need to wait as much.''} Both appreciated that the timer was customizable.

\subsubsection{\sysname{} Kept Users Informed through Detailed Feedback}
All participants valued the real-time transcription and feedback provided by \sysname{}. P2 appreciated seeing how the system recognized his commands, noting that transcription errors helped him prepare alternative approaches. P12 found the microphone indicator particularly helpful for understanding when he could and couldn't issue commands.

While the suggestions provided when commands couldn't be corrected proved useful, we noted that the LLM sometimes suggested commands based on syntactic viability rather than perceived user intent. For example, it might suggest a delete command when the user was attempting to insert text. Participants occasionally found these misaligned suggestions distracting.

\subsubsection{Perception of Time Delay}
Half of the participants noted the latency introduced by LLM processing and text-to-speech conversion. Particularly, the API response time depends on the internet connectivity and can take up to 2-3 seconds. P7 commented that while the delay was acceptable for occasional edits, it could become problematic during extended editing sessions requiring many consecutive commands.

%% file: source/sections/7_fine-tuning.tex
\section{Reducing Response Latency of \sysname{}}                                                                         
While \sysname{} improved user experience by allowing users to speak commands naturally and correcting the commands to the fixed-format, the time required for LLM API calls also introduced latency. In addition, depending on external APIs has cost implications and requires stable internet connectivity.

To further improve the system, we aimed to reduce response latency and run the model locally to eliminate costs and connectivity issues. Therefore, we fine-tuned a small language model, Gemma 3, with 270 million parameters, to convert users' commands to VUI format and served the fine-tuned model locally through Ollama\footnote{\url{https://ollama.com/}}. The fine-tuning process is informed by our formative study. We describe the process and the results in this section.

\paragraph{Dataset Generation for Fine-Tuning}

\begin{table*}[t!]
    \small
    \centering
    \begin{tabular}{p{3cm} | p{8cm} | p{4cm}}
         \textbf{Sample Type} & \textbf{Input} & \textbf{Expected Output} \\\toprule
         Exact command & select previous word | selection: apple & SELECT PREVIOUS WORD \\\midrule
         Natural utterance & can you please select the next word | selection: apple & SELECT NEXT WORD\\\midrule
         Swap \cmd{} & choose the word meeting | selection: & SELECT meeting\\\midrule
         Substitute \cmd{} & fix meeting | selection: & CORRECT meeting\\\midrule
         Substitute \cmd{}, Natural utterance & please add at home before that | selection: tonight & INSERT at home BEFORE tonight\\\midrule
         Missing arg & insert before apple pie | selection: & ASK: What should I insert before apple pie?\\\midrule
         Missing arg (follow-up clarification) &  insert before apple pie | selection: | CLARIFICATION QUESTION: What should I insert before apple pie? | CLARIFICATION: in the morning & INSERT in the morning BEFORE apple pie\\
        \bottomrule
    \end{tabular}
    \caption{Sample input and expected output from the synthetic dataset.}
    \Description{Sample input and expected output from the synthetic dataset.}
    \label{tab:synth-data}
\end{table*}

We generated synthetic fine-tuning data using a large language model, creating separate datasets for training, validation, and testing. We started with the commands collected during our prior studies and created incorrect and correct command pairs. We then provided these pairs to Claude Sonnet 4.5 along with the command templates (Figure~\ref{fig:cmnd-template}), example commands (Figure~\ref{fig:cmnd-comb}), and the incorrect command categories (Table~\ref{tab:incorrect-commands}), prompting it to generate samples for each command pair, both correct-to-correct and incorrect-to-correct mappings. We also generated samples where incorrect commands have missing information, requiring further clarification from users. 

We reviewed the generated samples, corrected any discrepancies, and provided the corrected examples back to Claude to refine the command correction process. We iterated this review-and-correction cycle until the generated samples consistently matched our quality standards. We then prompted the model to generate three synthetic datasets by varying parameters, substituting synonymous commands, and adding natural phrases (e.g., ``can you please...'', ``I want to...'') observed during our prior studies. Our final dataset consisted of 1,000 training samples, 400 validation samples, and 150 test samples. 

Each sample contains a fixed task (passed as the system prompt: \textit{Convert the following natural language command to the correct voice control command format.}), an input, and an expected output. For commands that are correct or contain all required information, the input includes the utterance and the current selection (which we maintain and update following each select command), and the expected output is the correct command expected by Voice Control. For commands with missing required information, the expected output is a follow-up question, which is concatenated to the input when the user responds. Example inputs and expected outputs are outlined in Table~\ref{tab:synth-data}.

\begin{figure*}[t!]
    \centering
    \begin{subfigure}{0.49\linewidth}
        \centering
        \includegraphics[width=\linewidth]{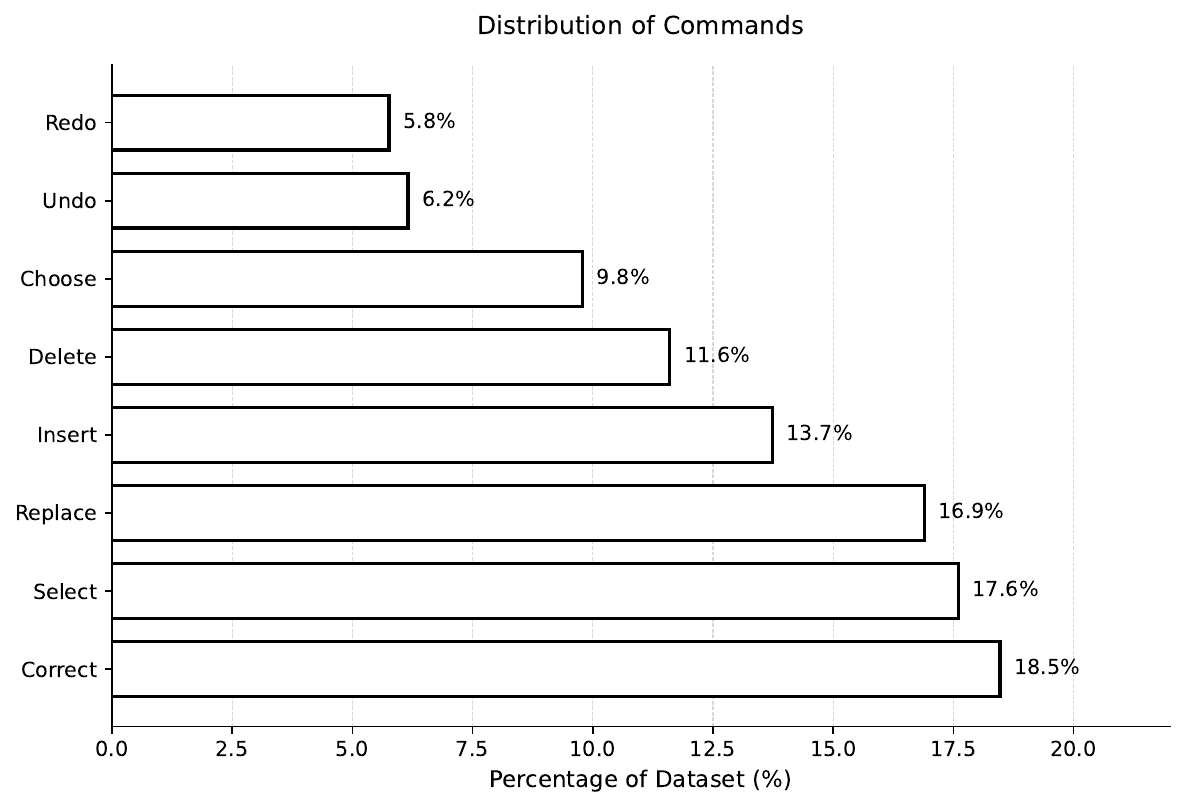}
        \caption{}
        \label{fig:cmd-dist}
    \end{subfigure}
    \begin{subfigure}{.49\linewidth}
        \centering
        \includegraphics[width=\linewidth]{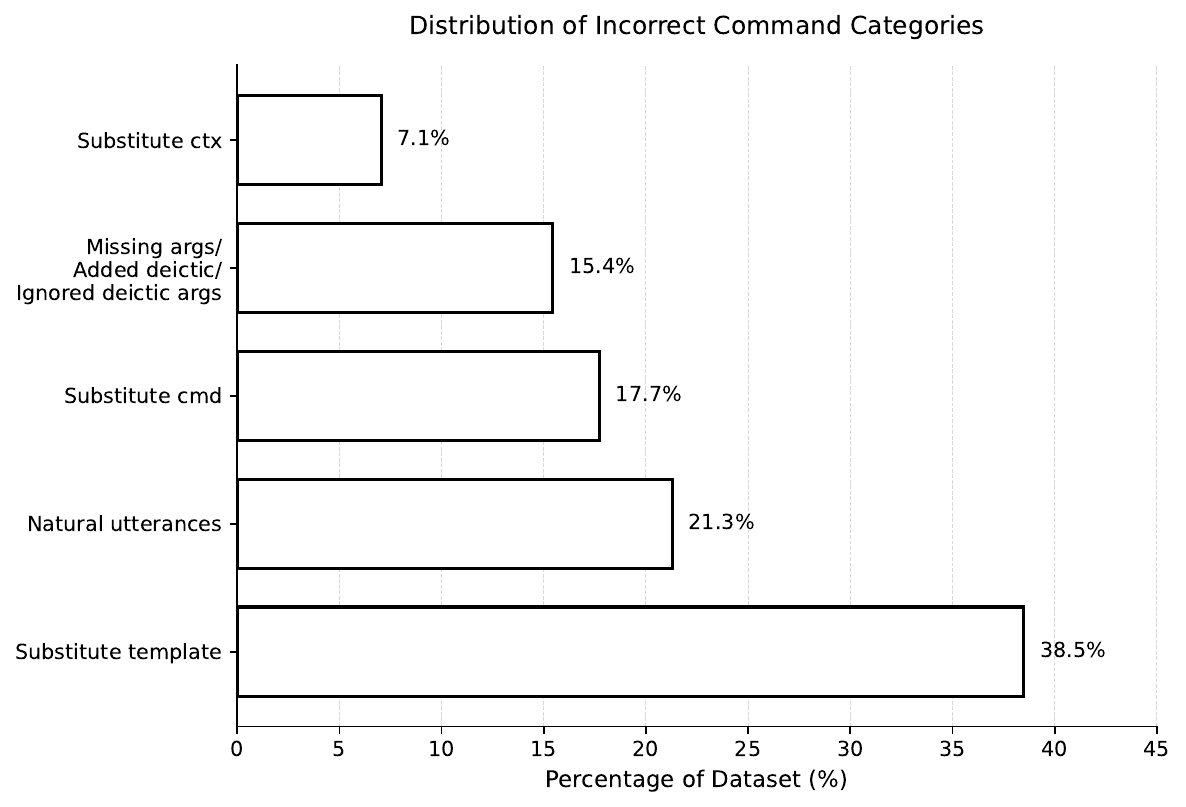}
        \caption{}
        \label{fig:error-dist}
    \end{subfigure}	
    \caption{Distribution of (a) text correction commands and (b) incorrect command categories in the synthetic dataset.}
    \Description{Distribution of (a) text correction commands and (b) incorrect command categories in the synthetic dataset.}
    \label{fig:data-dist}
\vspace{-1em}
\end{figure*}

Our synthetic dataset covers all six valid command structure combinations from Figure~\ref{fig:cmnd-comb}, with balanced representation across 8 command operations from Table~\ref{tab:voice_commands}, which is shown in Figure~\ref{fig:cmd-dist} (\texttt{\small correct}: 18.5\%, \texttt{\small select}: 17.6\%, \texttt{\small replace}: 16.9\%, \texttt{\small insert}: 13.7\%, \texttt{\small delete}: 11.6\%, \texttt{\small choose}: 9.8\%, \texttt{\small undo}: 6.2\%, \texttt{\small redo}: 5.8\%). The dataset captures all error categories from Table~\ref{tab:incorrect-commands}, which is shown in Fig.~\ref{fig:error-dist}: substitute template (38.5\%), natural utterances (21.3\%), substitute cmd (17.7\%), missing args/added deictic/ignored deictic args (15.4\%), and substitute ctx (7.1\%). We incorporated 50+ synonym variations informed by our formative study. The training samples represent 195 distinct correct commands, with multiple natural language variations for each (5.1 variations per command on average). This high variation density ensures the model learns to adapt diverse user inputs to correct command syntax rather than memorizing fixed patterns. The dataset balances commands with (29.2\%) and without (70.8\%) selection context, reflecting the full range of complexity observed in our formative study.

\paragraph{Fine-tuning Process}
We fine-tuned the Gemma 3 270M instruction-tuned model using parameter-efficient fine-tuning with LoRA~\cite{hu2022lora} (r = 128). Training was performed on an NVIDIA RTX A6000 GPU using the Unsloth framework~\cite{unsloth} for efficient fine-tuning. We used the SFTTrainer with AdamW optimizer, with approximately 10\% of the model's parameters being trainable. Validation loss was calculated after each epoch, and the model was saved accordingly.

\paragraph{Evaluation}
We evaluated the model before and after fine-tuning on the test set using exact match accuracy and ROUGE-L score~\cite{lin2004rouge} for semantic similarity. 
Before fine-tuning, the model achieved an exact match accuracy of 8.05\% and a ROUGE-L score of 0.2791 with zero-shot prompting. With 5-shot prompting, the accuracy and the ROUGE-L score slightly increased to 11.82\% and 0.3145, respectively.

After fine-tuning, the model achieved 90.6\% exact match accuracy and a ROUGE-L score of 0.9587. Upon closer examination of mismatches, we observed that the model occasionally asked clarifying questions in rare cases where commands were ambiguous. One such example is ``transform work'' where ``transform'' is used as a synonym of ``replace'' and the expected output asks ``what should I replace work with?'' The model instead asked ``what should I correct?'' However, this example represents an edge case with an uncommon synonym that never appeared in our user studies, suggesting the model performs well on realistic user inputs.

We converted the fine-tuned model into GGUF format and served it using Ollama to measure inference performance on a MacBook Pro. The model responded within 200 ms, with 95\% of responses completing in 130-150 ms. This demonstrates that the fine-tuned model enables real-time interaction on local devices while eliminating API latency and costs.

Overall, we found that smaller models fine-tuned for specific tasks informed by user studies can be accurate, fast, and effective for local on-device inference, providing greater flexibility for voice command systems. We plan to release the data and code in the future.

%% file: source/sections/8_discussion.tex
\section{Discussion}
Our studies revealed consistent patterns in user behavior. Participants instinctively selected targets before performing editing operations, even though this approach required multiple disambiguations. When legacy systems discard the context of earlier selections, they miss an opportunity to align with users' mental models and create frustration. While we partially mitigated this issue by extending select and replace commands, the black-box nature of these systems prevented a complete solution.

Our participants struggled with four-component commands, which demanded significant attention. Providing sufficient time and reducing commands to two components substantially improved their experience. We also observed that participants unconsciously reverted to synonyms even after mastering the correct commands. This finding emphasizes the importance of supporting minor variations in command phrasing while preserving semantic intent.

\subsection{Supporting Human-AI Interaction in VUIs}

Our work with \sysname{} yields insights that extend beyond adapting legacy VUI systems to inform broader questions in human-AI (HAI) interaction, particularly regarding how AI can mediate between natural human communication and rigid system constraints.

\subsubsection{Balancing Human Planning Time with AI Response Speed}
A fundamental issue in voice-based human-AI interaction is providing users sufficient time for cognitive processing while maintaining real-time responsiveness. Users need time to plan multi-component utterances, retrieve arguments from memory, or visually locate information, yet they expect immediate system feedback characteristic of natural conversation. AI-mediated voice systems must balance these opposing needs while providing distinguishable signals: indicators showing when the system is listening versus processing, real-time transcription maintaining users in the interaction loop, and minimized AI latency that preserves the conversational flow. This temporal design challenge is particularly acute for shimming layers augmenting legacy systems where users have internalized expectations of fast response times. Therefore, any AI mediation must feel seamless rather than introducing perceptible lag that disrupts the natural dialogue experience.

\subsubsection{Context Retention and Progressive Collaboration to Minimize Repetition}

Speech is inherently conversational and context-dependent, yet rigid command structures that require repetition and discard context undermine voice as a natural interaction modality. AI systems should function as conversational partners that maintain context across turns and scaffold incomplete input through targeted clarification rather than forcing users to restart. When users provide partial commands, AI should retain available information and ask minimal clarifying questions that allow users to continue their train of thought rather than reconstruct entire commands. This context awareness extends to leveraging information from immediate prior turns --- understanding deictic references, implicit arguments, and incomplete commands that characterize natural speech. Overall, AI must adapt to human conversational norms rather than requiring humans to adapt to system constraints, which can lead to a fluent HAI interaction experience.

\subsubsection{Small Specialized Models for Real-Time Structured Tasks}

Our findings indicate that fixed-format voice commands create conflicting requirements that small, specialized AI models can uniquely resolve. Users cannot reliably produce syntactically perfect commands spontaneously --- natural speech contains variations, false starts, and fillers --- yet the real-time responsiveness that makes voice feel natural precludes reliance on large general-purpose models that introduce latencies and external dependencies. For systems with predictable command structures and bounded vocabularies, local fine-tuned models can handle natural variation (synonyms, reordering, disfluencies) while maintaining faster responses. Therefore, when interactions involve mapping natural variation to known formal structures within bounded domains, small specialized models can provide reasonable accuracy, real-time performance, privacy through edge deployment, cost efficiency, and personalization opportunities compared to large general models, enabling voice interfaces that feel both natural and responsive.

\subsection{Extending VUI Systems through \sysname{}}

Emerging agentic interaction frameworks like the Model Context Protocol (MCP)~\cite{anthropic2024mcp} represent promising directions for AI-system integration. However, they face fundamental limitations in legacy computer control scenarios. MCP-based approaches require direct API access to application DOM structures and system internals --- capabilities that are increasingly restricted in modern operating systems like macOS for security reasons. Legacy VUI systems like Voice Control bypass these restrictions through deep OS-level integration with accessibility APIs, providing comprehensive system control that external frameworks cannot replicate easily. Our shimming layer preserves this valuable low-level access of VUIs while adding an adaptability layer on top, offering a practical middle ground between complete system replacement and acceptance of rigid command structures.

While we focused on text correction commands, our approach of decomposing voice commands into discrete components and analyzing how users combine these components to match command templates applies broadly to other commands supported by fixed-format VUIs, such as computer control commands (\texttt{\small click <icon>}, \texttt{\small open <application>}). Our methodology of breaking down seemingly simple commands into granular components and determining their interrelationships and usability can extend to any VUI command category. Commands with certain characteristics are particularly prone to failure: those requiring many components uttered sequentially, those requiring users to search for or identify information mid-speech, and those referencing contextual information increase cognitive load and failure rates. In such cases, breaking commands into smaller steps, minimizing required components, reducing component variations (e.g., variable component ordering or differing template lengths), and standardizing valid command structures can improve predictability. Therefore, for any VUI system, identifying a comprehensive command template, enumerating valid component combinations, analyzing component interconnectivity, and determining how each component is acquired during utterance can reveal complexity patterns and guide better voice command design.

Our fine-tuning of a local LLM keeps system commands intact while supporting minor command variations to reduce cognitive demand and command retries. This approach generalizes to other VUI systems by generating synthetic datasets with command variations and fine-tuning local models for those systems. The fine-tuning methodology also enables personalization, as not all commands need adaptation. Users could select their most frequently used commands and provide example variations, allowing an LLM to generate a personalized synthetic dataset that enables our shim layer to adapt variations most relevant to individual workflows, leading to better usability.

Beyond command adaptation, our \sysname{} architecture can address other VUI limitations. For multilingual support, our approach could help users interact with English-only VUIs using their preferred languages. Recent LLMs demonstrate strong translation capabilities~\cite{zhang2023prompting, robinson2023chatgpt}, potentially allowing users to speak commands in their native language before \sysname{} translates them into English and relays them to the legacy system. Similarly, this approach could improve VUI performance in noisy environments by preprocessing commands to separate the signal from noise~\cite{xu2014regression, wang2018supervised}. The command adapter could filter background interference and extract critical command components before sending a clean, structured command to the underlying system.

\subsection{Privacy, Trust, and Transparency in AI-Mediated VUI Interaction}

Privacy, trust, and transparency have been longstanding concerns for VUIs, particularly with the integration of AI~\cite{lau2018alexa, abdi2021privacy}. When voice commands are processed through cloud-based LLM APIs, sensitive user data, including document content, personal communications, and contextual information, is transmitted to external servers where it may be retained, aggregated, or used for model training without explicit user consent~\cite{malkin2019privacy}. This practice constitutes a potential breach of user privacy, and the opacity of data handling practices by API providers undermines transparency~\cite{apthorpe2018keeping}. Prior research has shown that VUI users are often unaware of how their data is collected and used, placing trust in systems to protect their privacy. However, users with heightened security concerns frequently express reluctance to adopt voice interfaces due to these privacy risks~\cite{lau2018alexa}.

While our initial \sysname{} implementation relied on a cloud-based LLM API and inherited these concerns, we subsequently leveraged insights and data from our formative study to fine-tune a small local model. This on-device deployment fundamentally addresses privacy concerns by ensuring all command processing occurs locally without external data transmission, aligning with recommendations for privacy-preserving edge-based AI systems~\cite{xu2021edge}. Beyond privacy, our design incorporates transparency and trust-building mechanisms. \sysname{} provides real-time visual feedback of detected speech transcription, seeks clarifying questions when commands contain incomplete information, and offers structured command suggestions rather than executing potentially incorrect transformations. In addition, our shim layer functions as an intermediary while preserving the original VUI's features and feedback mechanisms, ensuring that users retain their familiar interaction experience with the underlying system. These features help users develop accurate mental models of system capabilities and limitations, which prior work has identified as essential for appropriate trust calibration in AI-mediated interactions~\cite{amershi2019guidelines, ehsan2021expanding}.

\subsection{Limitations and Future Work}
Our work has several limitations. First, using text-to-speech to communicate with the VUI through the virtual channel introduces two challenges: TTS adds latency as commands must be played again for the interface, and VUIs may struggle to recognize robotic TTS output, occasionally resulting in discarded commands, as we observed during our study. For VUIs supporting direct text input, bypassing TTS could improve reliability.

Second, although our study included participants with varying accents, all were fluent English speakers. Consequently, our findings may not fully reflect the experiences of non-fluent English speakers or users with speech impairments. The cognitive load and task completion strategies of such users when forming voice commands may differ substantially, representing an important direction for future research toward a more comprehensive understanding of VUI command usability.

Third, this work primarily investigates VUIs supporting fixed-format commands. Our analysis of four commercial VUIs and their command structures informed our fully quantified command template, which we reinforced through our formative study. While we believe this template represents command formats typical of VUIs for personal computing devices, it may not encompass the entire VUI design space (e.g., smart home controls, in-car systems, or industrial applications).

Finally, our system maps one voice command to one VUI command, which is a common practice for legacy VUIs. However, with natural commands, it is possible to combine multiple voice commands to allow users more flexibility. Additionally, we focused on text correction commands while VUIs, such as Voice Control, have commands to perform all sorts of different UI control tasks. In the future, we aim to explore such commands to allow users control their devices through natural language commands utilizing the technical capabilities of legacy VUIs underneath.

%% file: source/sections/9_conclusion.tex
\section{Conclusion}

Our work demonstrates how a shimming layer enhances legacy VUI usability by treating them as black boxes. By addressing rigid command formats, restrictive timeouts, and insufficient feedback, our proposed \sysname{} translates natural voice commands into the precise syntax required by legacy VUI systems. Our evaluation showed substantial improvements: reducing command failures, requiring fewer commands per task, and significantly reducing cognitive and temporal demands. Furthermore, our fine-tuned Gemma 3 270M model enables deployment with over 90\% accuracy and 200 ms response time when served locally, eliminating dependence on third-party APIs while enabling real-time interaction on edge devices. This approach offers a practical way to improve voice interactions without replacing existing infrastructure, illustrating how modern LLM techniques can support human-AI interaction.